\documentclass[twocolumn,amsmath,amssymb,showpacs,superscriptaddress]{revtex4-1}
\usepackage{graphicx}
\usepackage{epstopdf}
\usepackage{dcolumn}
\usepackage{bm}
\usepackage{amsmath}
\usepackage{amsfonts}
\usepackage{amssymb}
\usepackage{mathrsfs}
\usepackage{color}
\usepackage{comment}
\newcommand{\pp}{{\prime\prime}}
\makeatletter

\begin{document}
\title{Rheology of dilute cohesive granular gases}
\author{Satoshi Takada}
\email[]{takada@eri.u-tokyo.ac.jp}
\affiliation{Earthquake Research Institute, The University of Tokyo, 1-1-1, Yayoi, Bunkyo-ku, Tokyo 113-0032, Japan}
\affiliation{Department of Physics, Kyoto University, Kitashirakawa Oiwakecho, Sakyo-ku, Kyoto 606-8502, Japan}
\author{Hisao Hayakawa}
\affiliation{Yukawa Institute for Theoretical Physics, Kyoto University, Kitashirakawa Oiwakecho, Sakyo-ku, Kyoto 606-8502, Japan}
\date{\today}

\begin{abstract}
Rheology of a dilute cohesive granular gas is theoretically and numerically studied.
The flow curve between the shear viscosity and the shear rate is derived from the inelastic Boltzmann equation for particles having square-well potentials in a simple shear flow.
It is found that (i) the stable uniformly sheared state only exists above a critical shear rate and (ii) the viscosity in the uniformly sheared flow is almost identical to that for uniformly sheared flow of hard core granular particles. 
Below the critical shear rate, clusters grow with time, in which the viscosity can be approximated by that for the hard-core fluids if we replace the diameter of the particle by the mean diameter of clusters.
\end{abstract}
\maketitle

%%%%%%%%%%%%%%%%%%%%%%%%%%%%%%
\section{Introduction}
Granular materials, having dissipative interactions between particles, are ubiquitous in daily life as unusual solids, liquids and
gases \cite{Jaeger1996}.
It is important to know rheological properties of granular flows to control the granular materials \cite{Jop2006,Boyer2011,MiDi2004,Silbert2001,Jenkins1985a, Jenkins1985b, Lutsko2005, Saitoh2007, Gnoli2016}.
The rheological properties of granular flows strongly depend on their densities ranging from dilute gases to the jammed solids.
When we focus on the rheology of granular flows for the density below the volume fraction $\varphi<0.5$, the description in terms of the
Boltzmann-Enskog equation gives quantitatively correct results \cite{Brey1998, Garzo1999, Brilliantov, Mitarai2007, Chialvo2013, Hayakawa2017}, while the appropriate theory for the description of denser flow is still controversial \cite{Suzuki2015, DeGiuli2015}.

So far, most of previous studies on dry granular flows assume that the interactions between grains can be described by repulsive and
dissipative forces. 
Attractive interactions, however, are not negligible for fine powders and wet granular particles \cite{Castellanos2005a, Herminghaus2013, Herminghaus2005b, Mitarai2006}.
The origins of such cohesive forces are, respectively, van der Waals force for fine powders and capillary force for wet granular particles.
Such attractive forces cause the liquid-gas phase transition and the clustering instability as well as the enhancement of the jamming
transition \cite{Rahbari2010,Chaudhuri2012,Gu2014,Singh2014,Takada2014,Saitoh2015,Irani2014,Irani2016}.
Therefore, to know the rheology of flows consisting of cohesive granular particles is important not only for engineers but also for
physicists.

In our previous paper \cite{Takada2016}, we have developed the systematic kinetic theory of freely cooling dilute cohesive granular particles in terms of the inelastic Boltzmann equation for particles having square-well potentials.
Nevertheless, we still need to analyze the rheology of cohesive granular particles under a simple shear flow, because (i) we are interested in a nonequilibrium steady state under the balance between an external force such as shear and the energy dissipation due to inelastic collisions, and (ii) the viscosity of granular flows under a simple shear differs from that for freely cooling granular gases \cite{Santos2004, Santos2007}.
Moreover, we have to consider the contribution of clustering caused by the attractive interaction between grains to the rheology
systematically.

In this paper, we try to clarify the rheological properties of dilute granular gases having an attractive interaction described by the square-well potential.
The organization of this paper is as follows.
In the next section, we explain the setup and the results of the event-driven simulation by DynamO \cite{Bannerman2011}.
In Sec.\ \ref{sec:kinetic_theory}, we derive the shear viscosity in terms of the Boltzmann equation to compare the results with those from the simulation.
We also briefly explain the result of the linear stability analysis.
In Sec.\ \ref{sec:summary}, we summarize our results.
Technical details are described in Appendices \ref{sec:LEbc}--\ref{sec:critical_temp}.

%%%%%%%%%%%%%%%%%%%%%%%%%

%%%%%%%%%%%%%%%%%%%%%%%%%%%%%%
\section{Molecular dynamics simulation under a simple shear}\label{sec:simulation}
In this section, we explain our model and the setup of our event-driven simulation in terms of DynamO \cite{Bannerman2011} for dilute cohesive granular gases under a uniform shear in Sec.\ \ref{sec:setup}.
We present the results of our simulation in Sec.\ \ref{sec:MD_results}.

\subsection{Our model}\label{sec:setup}
We consider a collection of monodisperse particles in which the mass and the diameter are, respectively, given by $m$ and $d$.
We assume that the interaction between particles is described by the square-well potential
\begin{equation}
	U(r)=
	\begin{cases}
	\infty & (r\le d)\\
	-\varepsilon & (d<r\le \lambda d)\\
	0 & (r>\lambda d)
	\end{cases},\label{eq:SW}
\end{equation}
where $r$, $\varepsilon$ and $\lambda$ are the distance between particles, the well depth and the ratio of the whole potential range to the hard-core repulsive range, respectively.
We assume that each collision is inelastic when two particles collide at $r=d$, and collisions are elastic otherwise.
Here, the inelasticity is characterized by the restitution coefficient $e$, which is the ratio of the post-collisional relative normal speed to the pre-collisional one.
Note that the detailed expressions of collision processes by this potential are presented in Ref.\ \cite{Takada2016}.
We also note that we are mainly interested in nearly elastic cases, i.e., $e\lesssim 1$ because the applicability of the kinetic theory for cohesive granular gases is limited in this region \cite{Takada2016}.
A simple shear flow characterized by the shear rate $\dot\gamma$ is applied in $x$-direction under the Lees-Edwards boundary condition \cite{Lees1972}.
(We show the results under the flat boundary condition and to clarify the artifacts caused by the periodic boundary condition
 in Appendix \ref{sec:LEbc}.)
The time evolutions of the position $\bm{r}_i$ and the velocity $\bm{v}_i$ of $i$-th particle are updated by the event-driven simulation for hard-core particles.
We mainly simulate the systems of $N=1,372$ particles in a cubic box, whose size is $L=41.6d$.
We also simulate the system of $N=10,976$ particles in a cubic box corresponding to $L=83.1d$ to check finite size effects.
Throughout this paper, we fix the packing fraction as $\varphi=N(\pi d^3/6)/L^3=0.01\ll 1$, the inelasticity $1-e=0.01, 0.1$, and $0.3$, and the ratio characterizing the potential well $\lambda=1.5$.
We measure various quantities by changing the dimensionless shear rate $\dot\gamma^*\equiv \dot\gamma \sqrt{md^2/\varepsilon}$.
We show the results for $1-e=0.01$ in the main text and present the results for $1-e=0.1$ and $0.3$ in Appendix \ref{sec:e_change} to clarify the role of inelasticity.

\subsection{Results}\label{sec:MD_results}

\begin{figure}[htbp]
	\begin{center}
		\includegraphics[width=80mm]{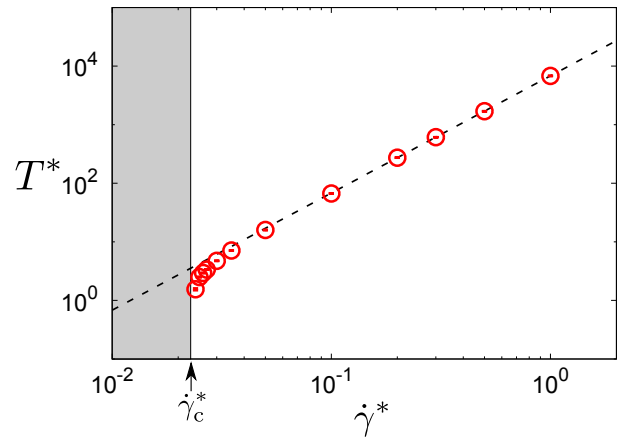}
	\end{center}
	\caption{We present the relationship between the temperature and the shear rate for $e=0.99$ and $\lambda=1.5$ (open circles).
	The dashed line expresses Bagnoldian expression Eq.\ (\ref{eq:T_Bagnold}).
	The shaded region represents the absence of steady state.}
	\label{fig:temp_sim}
\end{figure}

\begin{figure}[htbp]
	\begin{center}
		\includegraphics[width=80mm]{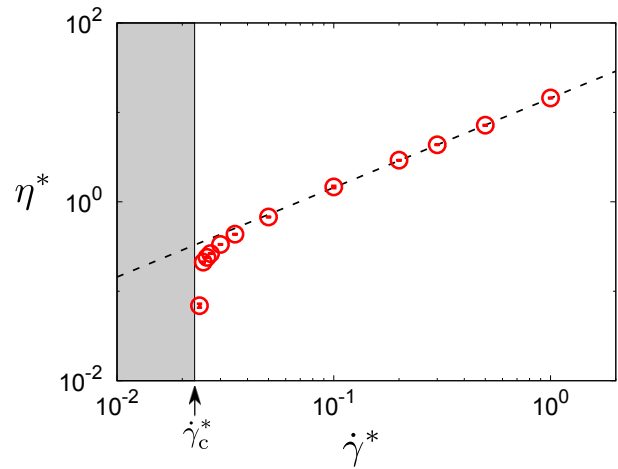}
	\end{center}
	\caption{We present the shear rate dependence of the shear viscosity for $e=0.99$ and $\lambda=1.5$ (open circles), where dashed line represents Bagnoldian temperature Eq.\ (\ref{eq:eta_Bagnold}).
	The shaded area expresses the region which does not have any steady state.}
	\label{fig:steady_gamma_eta_sim}
\end{figure}

Let us present the results of our MD.
Figures \ref{fig:temp_sim} and \ref{fig:steady_gamma_eta_sim} exhibit the results of the dimensionless kinetic or granular temperature $T^*\equiv T/\varepsilon$ and the shear viscosity $\eta^*\equiv \eta d^2/\sqrt{m\varepsilon}$ against the dimensionless shear rate $\dot\gamma^*$, respectively, in steady states above the critical shear rate for $e=0.99$.
Here, the shear viscosity is defined by $\eta=-P^k_{xy}/\dot\gamma$, where the time-averaged kinetic part of the stress tensor $\overleftrightarrow{P}$ is expressed as \cite{Alam2003, Bannerman2011}:
\begin{equation}
	\overleftrightarrow{P}^k=\frac{1}{L^3} \sum_{i=1}^N m \bm{V}_i \bm{V}_i.\label{eq:P_measurement}
\end{equation}
Here, $\bm{V}_i=\bm{v}_i -\dot\gamma y_i \hat{\bm{e}}_x$ is the peculiar velocity of $i$-th particle with the unit vector in $x$-direction $\hat{\bm{e}}_x$.
Note that the stress tensor for dilute gases should be dominated by the kinetic part even if clustering takes place (see Appendix \ref{sec:ratio_Pk_P}).
To obtain stabilized data, the stress is time-averaged during the dimensionless time interval $100$.

%%%%%%%%%%%%%%%%%%%%%%%%%%%%%%
%%%%%%%%%%%%%%%%%%%%%%%%%%%%%%
%%%%%%%%%%%%%%%%%%%%%%%%%%%%%%
%%%%%%%%%%%%%%%%%%%%%%%%%%%%%%

Figures \ref{fig:temp_sim} and \ref{fig:steady_gamma_eta_sim} indicate the existence of the critical shear rate $\dot\gamma_{\rm c}^*(\equiv \dot\gamma_{\rm c} \sqrt{\varepsilon/md^2}=0.023)$ above which there exist steady states.
We also plot steady Bagnoldian expressions for the kinetic temperature and the viscosity
\begin{align}
T_{\rm B}^* &= \frac{5\pi (2+e)}{432(1-e)(1+e)^2(3-e)^2}\frac{1}{\varphi^2}md^2\dot\gamma^2/\varepsilon,\label{eq:T_Bagnold}\\
\eta_{\rm B}^* &= \frac{5(2+e)}{72(1+e)^2(3-e)^3}\sqrt{\frac{5(2+e)}{3(1-e)}}\frac{1}{\varphi}\sqrt{\frac{md^2}{\varepsilon}}\dot\gamma,\label{eq:eta_Bagnold}
\end{align}
of hard-core dilute granular gases in Figs.\ \ref{fig:temp_sim} and \ref{fig:steady_gamma_eta_sim} \cite{Santos2004}.
It is remarkable that Eqs.\ (\ref{eq:T_Bagnold}) and (\ref{eq:eta_Bagnold}) give precise results for $\dot\gamma >\dot\gamma_{\rm c}$ except for the region in the vicinity of $\dot\gamma_{\rm c}$.
These results can be understood as follows:
When the shear rate is sufficiently larger than the critical one, the temperature determined by the energy balance is also larger than the well depth, where the attractive force is negligible.
This is the reason why the flow curve reduces to Bagnoldian expressions in the high shear regime.

%%%%%%%%%%%%%%%%%%%%%%%%%%%%%%
\begin{figure*}[htbp]
	\begin{center}
		\includegraphics[width=180mm]{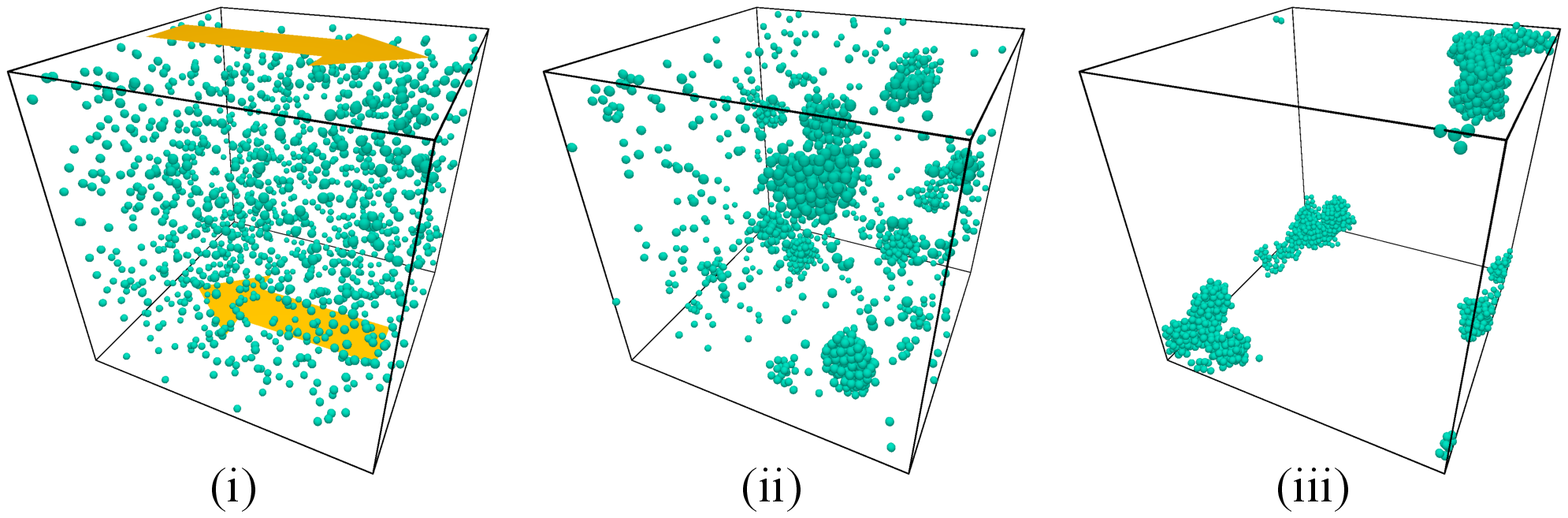}
	\end{center}
	\caption{Typical snapshots of the system for $\dot\gamma^*=0.01$ at (i) $t^*=0$, (ii) $190$, and (iii) $315$, where the arrow indicates the direction of the shear.}
	\label{fig:cluster}
\end{figure*}
%%%%%%%%%%%%%%%%%%%%%%%%%%%%%%
%%%%%%%%%%%%%%%%%%%%%%%%%%%%%%
\begin{figure}[htbp]
	\begin{center}
		\includegraphics[width=80mm]{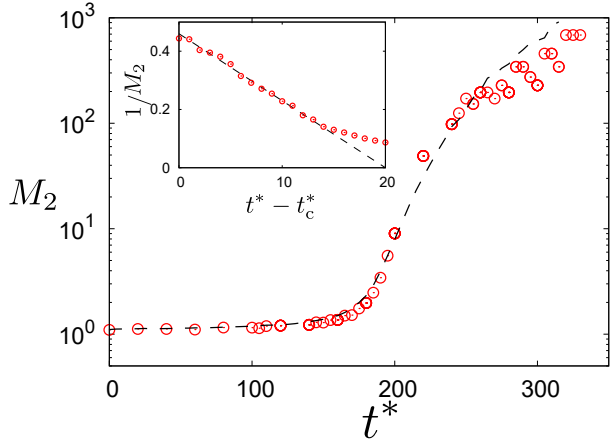}
	\end{center}
	\caption{The time evolution of the mean cluster size $M_2$ for $\dot\gamma^*=0.01$ and $N=1,372$ (open circles) as well as the data for $\dot\gamma^*=0.01$ and $N=10,976$ (dashed line).
	The inset shows the time evolution after $t_{\rm c}^*=180$ for $\dot\gamma^*=0.01$, where the dashed line is a fitting function (\ref{eq:Ncl_evol}).}
	\label{fig:cluster_evol}
\end{figure}
%%%%%%%%%%%%%%%%%%%%%%%%%%%%%%

For small shear rate, there does not exist any steady state.
Figure \ref{fig:cluster} is the time evolution of clustering process observed in our MD for $\dot{\gamma}^*=0.01$.
The time evolution of the mean cluster size is plotted in Fig.\ \ref{fig:cluster_evol}, in which two particles belongs to a same cluster when the distance between them is less than $\lambda d$.
The mean cluster size is almost unity for large shear rate ($\dot\gamma>\dot\gamma_{\rm c}$), while it drastically increase after $t^*=180$ for small shear rate ($\dot\gamma < \dot\gamma_{\rm c}$), where we have introduced the dimensionless time $t^*\equiv t/\sqrt{md^2/\varepsilon}$ and the second moment of the cluster size $M_2=\sum_{k=1}^\infty k^2 c_k$ with the size distribution $c_k$ of the size $k$. 
It should be noted that the growth rate of the mean cluster size becomes smaller after $t^*=250$.
It should be noted that $M_2$ can be regarded as the mean cluster size because $M_1=\sum_{k=1}^\infty k c_k$ is always equal to the unity.
Note that this tendency seems to be insensitive to the system size from the comparison of the results of $N=1,372$ with those of $N=10,976$.
The inset of Fig.\ \ref{fig:cluster_evol} tries to compare the cluster growth with
\begin{equation}
	\frac{1}{M_2} =\alpha_1 - \alpha_2 (t-t_{\rm c}),\label{eq:Ncl_evol}
\end{equation}
for $t>t_{\rm c}=180\sqrt{md^2/\varepsilon}$. 
Here, we have introduced the fitting parameters $\alpha_1$ and $\alpha_2$, where the fitting range is $0\le t^*-t_{\rm c}^* \le 15$.
The justification of this fitting curve will be discussed later.

%%%%%%%%%%%%%%%%%%%%%

%%%%%%%%%%%%%%%%%%%%%%%%%%%%%%
\begin{figure}[htbp]
	\begin{center}
		\includegraphics[width=80mm]{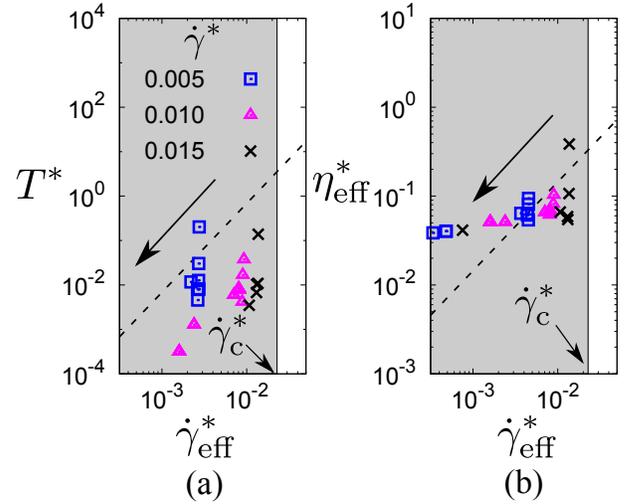}
	\end{center}
	\caption{Plots of (a) the temperature and (b) the effective shear viscosity Eq.\ (\ref{eq:eff_vis_cl}) against the effective shear rates Eq.\ (\ref{eq:gamma_vis_cl}) for several shear rates, respectively.
	The arrow indicates the time evolution, and all quantities decrease with time.}
	\label{fig:effective}
\end{figure}
%%%%%%%%%%%%%%%%%%%%%%%%%%%%%%
Let us introduce the effective shear viscosity and the effective shear rate scaled by $M_2$.
We adopt two assumptions:
First, each cluster can be replaced by a sphere which has the identical size.
Next, we ignore the size distribution or size fluctuation of clusters.
From these assumptions, the diameter and the mass of the clusters are, respectively, given by $d_{\rm cl}=M_2^{1/3}d$ and $m_{\rm cl}=(d_{\rm cl}/d)^3m=M_2m$.
Therefore, we can introduce the effective shear viscosity and the effective shear rate 
\begin{align}
\eta_{\rm eff}&=\eta_{\rm c} M_2^{-1/6},\label{eq:eff_vis_cl}\\
\dot\gamma_{\rm eff}&=\dot\gamma_{\rm c}^* \sqrt{\frac{\varepsilon}{m_{\rm cl}d_{\rm cl}^2}}
	=\dot\gamma_{\rm c}M_2^{-5/6},\label{eq:gamma_vis_cl}
\end{align}
respectively, where $\eta_{\rm c}$ is the critical shear viscosity at $\dot\gamma_{\rm c}$.
We, respectively, plot the time evolution of the temperature versus $\dot\gamma_{\rm eff}$ introduced in Eq.\ (\ref{eq:gamma_vis_cl}) and the relationship between Eqs.\ (\ref{eq:eff_vis_cl}) and (\ref{eq:gamma_vis_cl}) in Figs.\ \ref{fig:effective}(a) and (b), where the dimensionless effective shear viscosity and shear rate are, respectively, introduced by $\eta_{\rm eff}^*\equiv\eta_{\rm eff} d^2/\sqrt{m\varepsilon}$ and $\dot\gamma_{\rm eff}^*\equiv \dot\gamma_{\rm eff}\sqrt{md^2/\varepsilon}$.
In our simulation, the initial temperature $T_0$ satisfies Eq.\ (\ref{eq:T_Bagnold}) at given $\dot\gamma^*$.
It is noteworthy that the effective viscosity is approximately represented by Bagnoldian expressions in the ranges $10^{-3}\lesssim \dot\gamma^* \lesssim \dot\gamma^*_{\rm c}$, though the flow curves evolve with time to access the origin.
Note that the kinetic temperature $T^*$ is not well approximated by such a crude treatment (see Fig.\ \ref{fig:effective}(a)).

%%%%%%%%%%%%%%%%%%%%%

\begin{figure}[htbp]
	\begin{center}
		\includegraphics[width=80mm]{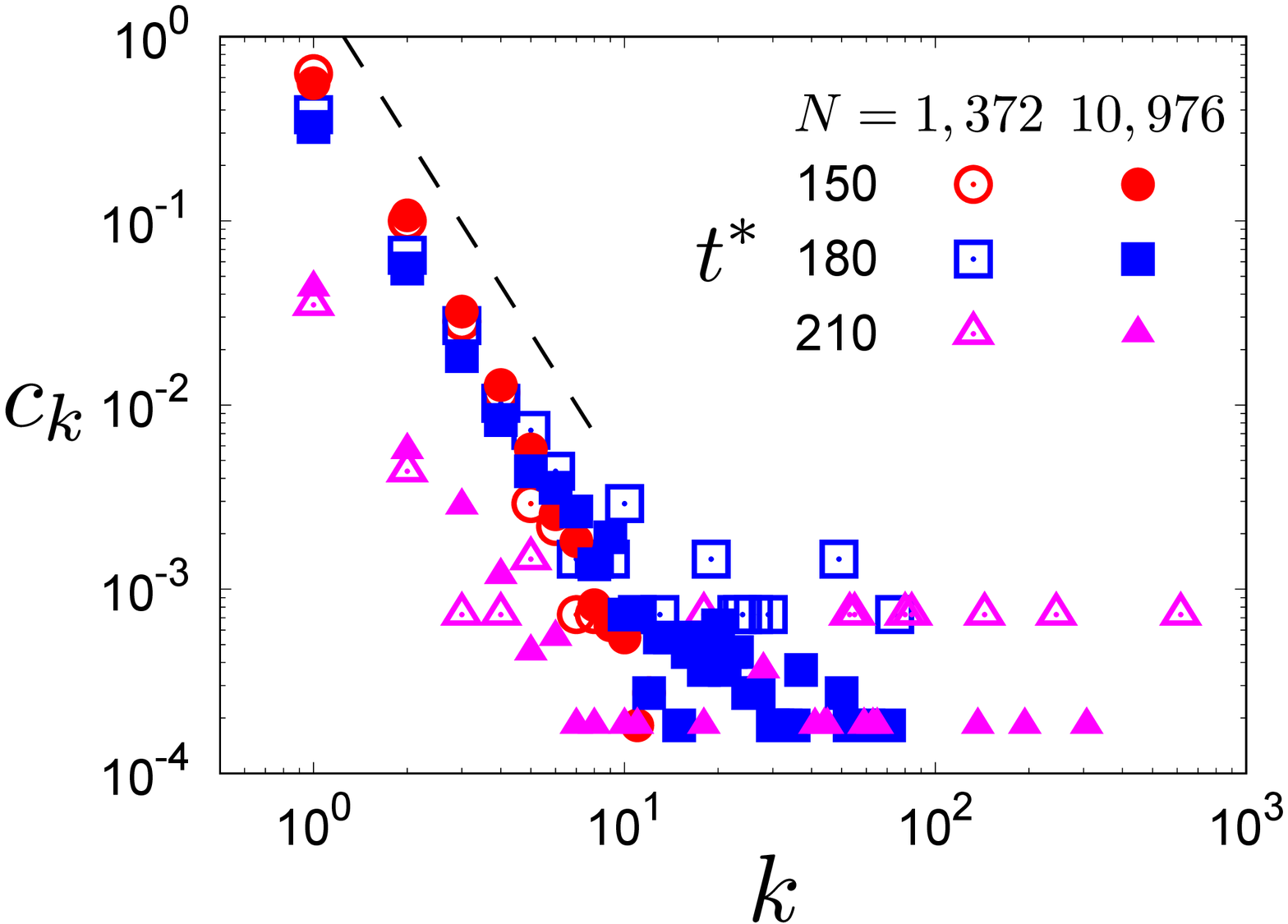}
	\end{center}
	\caption{The cluster size distribution at (i) $t^*=150$ (open circles), (ii) $180$ (open squares), and $210$ (open triangles) for $\dot\gamma=0.01$ and $N=1,372$ as well as the data for $\dot\gamma^*=0.01$ and $N=10,976$ (corresponding solid marks).
	The black dashed line represents $c_k\propto k^{-\beta}$ with $\beta\simeq 2.67$.}
	\label{fig:size_dist}
\end{figure}

We also investigate the cluster size distribution when the clustering proceeds.
For small clusters, $c_k$ can be well fitted by a power law as
\begin{equation}
	c_k \propto k^{-\beta}\label{eq:power_law}
\end{equation}
for $1\le k \le 10$, where the exponent is $\beta\simeq 2.67$ (see Fig.\ \ref{fig:size_dist}). 
This broad size distribution, though the cutoff size is not large, is incompatible with the assumption (ii) in which the sized distribution is negligible.
It should be noted that $\beta$ is almost independent of time in the range $180\le t^*\le 250$ and the system size as shown in Fig.\ \ref{fig:size_dist}, when the mean cluster size drastically increases.
We also note that this exponent is insensitive to the shear rate within the range $0.003\le \dot\gamma^*\le 0.015$ and the system size.

%%%%%%%%%%%%%%%%%%%%%%%%%%%%%%
%%%%%%%%%%%%%%%%%%%%%%%%%%%%%%

We discuss the evolution of cluster size described by Eqs.\ (\ref{eq:Ncl_evol}) and (\ref{eq:power_law}) in the unstable region.
This process might be explained by Smoluchowski's rate equation \cite{Ziff1980, Ziff1982, Hendriks1983}
\begin{equation}
	\frac{dc_k(t)}{dt}=\frac{1}{2}\sum_{i+j=k} K_{i,j}c_i(t)c_{j}(t)-c_k(t)\sum_{j=1}^\infty K_{k,j}c_j(t). \label{eq:Smoluchowski}
\end{equation}
Judging from Eqs.\ (\ref{eq:Ncl_evol}) and (\ref{eq:power_law}), we may use the corresponding coagulation kernel $K_{i,j}=K_0 ij$.
It is noted that this kernel can be applied to systems where all the elements are equally reactive in polymerization processes \cite{Ziff1980, Ziff1982, Hendriks1983}.
Because the time evolution of $M_2(t)$ can be explicitly solved as $M_2(t)=M_2(0)/(1-2M_2(0)t)$ \cite{Ziff1980}, the mean cluster size is given by
\begin{equation}
	M_2(t)=\frac{M_2(0)}{1-2M_2(0)K_0 t},\label{eq:N_cl_evol}
\end{equation}
which qualitatively agrees with Eq.\ (\ref{eq:Ncl_evol}).
We also note that the size distribution in the vicinity of the gelation satisfies $c_k\sim k^{-5/2}$ which is similar to Eq.\ (\ref{eq:power_law}).
At present, the applicability of Smoluchowski's equation (\ref{eq:Smoluchowski}) to our clustering process is not clear.
Further investigation along this line will be needed.

Before closing this section, let us briefly summarize the results for smaller $e$ such as $e=0.7$ and $0.9$ (see Appendix \ref{sec:e_change}).
Even if we are interested in moderately dissipative situations, the qualitative behavior is common, i.\ e.\ (i) Bagnoldian expressions can be used for highly sheared cases, (ii) there is a critical shear rate that the uniform state is unstable.
In particular, we should note that the the Bagnoldian expression for $\eta_{\rm eff}^*$ is still valid in clustering regime for $\dot\gamma_{\rm eff}^*>10^{-3}$.

%%%%%%%%%%%%%%%%%%%%%%%%%%%%%%%
%%%%%%%%%%%%%%%%%%%%%%%%%%%%%%%

\section{Kinetic theory}\label{sec:kinetic_theory}
In the previous section, we numerically found the existence of the critical shear rate $\dot\gamma_{\rm c}$, below which there is no steady state.
In this section, let us consider the Boltzmann equation for granular gases having the square-well potential Eq.~(\ref{eq:SW}) under a simple shear flow.
Although we have considered a shear driven by the Lees-Edwards boundary condition in the simulation, we consider a bulk shear in the treatment of the kinetic theory for simplicity.
Thus, we evaluate the steady observables in terms of the Boltzmann equation and compare the theoretical results with those obtained by the simulation in the previous section.
We also clarify what determines this critical shear rate.
Note that such a theoretical analysis is only possible for nearly elastic cases $e\lesssim 1$.

Let us begin with the Boltzmann equation \cite{Chapman}
\begin{equation}
\left(\frac{\partial}{\partial t}+\bm{v}_1\cdot \bm\nabla\right)f(\bm{r},\bm{v}_1,t)
= J(\bm{v}_1|f),\label{eq:Boltzmann}
\end{equation}
for a dilute gas consisting of particles interacting through the square-well potential in Eq.\ (\ref{eq:SW}),
where $J(\bm{v}_1|f)$ is the collision integral
\begin{align}
J(\bm{v}_1|f)&=\int d\bm{v}_2 \int d\hat{\bm{k}} \Theta(\min(\lambda,\mathfrak{N})-\tilde{b}) v_{12}\nonumber\\
	&\hspace{2em}\times\left[\mathscr{J} \sigma(\chi, v_{12}^\pp) f(\bm{r},\bm{v}_1^\pp,t)f(\bm{r},\bm{v}_2^\pp,t) \right.\nonumber\\
	&\hspace{3em}\left.-\sigma(\chi, v_{12}) f(\bm{r},\bm{v}_1,t)f(\bm{r},\bm{v}_2,t)\right]\nonumber\\
	&\hspace{2em}+\int d\bm{v}_2 \int d\hat{\bm{k}} \Theta(\tilde{b}-\min(\lambda,\mathfrak{N})) v_{12}\nonumber\\
	&\hspace{2em}\times \left[\sigma(\chi, v_{12}^\pp) f(\bm{r},\bm{v}_1^\pp,t)f(\bm{r},\bm{v}_2^\pp,t) \right.\nonumber\\
	&\hspace{3em}\left.- \sigma(\chi, v_{12}) f(\bm{r},\bm{v}_1,t)f(\bm{r},\bm{v}_2,t)\right].
\end{align}
Here, we have introduced the step function $\Theta(x)=1$ for $x\ge 0$ and $\Theta(x)=0$ otherwise,
the refractive index $\mathfrak{N}\equiv (1+4\varepsilon/mv_{12}^2)^{1/2}$ \cite{Landau, Goldshtein}, 
$\tilde{b}=b/d$, 
$v_{12}=|\bm{v}_{12}|=|\bm{v}_1-\bm{v}_2|$,
the Jacobian $\mathscr{J}$ of the transformation between pre-collisional velocities ($\bm{v}_1^{\prime\prime}$, $\bm{v}_2^{\prime\prime}$) and the post-collisional velocities ($\bm{v}_1$, $\bm{v}_2$), and the collision cross section $\sigma(\chi,v_{12})$ between particles $1$ and $2$ at the scattering angle $\chi$.
For the square-well potential, the relationship between $(\bm{v}_1^\pp, \bm{v}_2^\pp)$ and $(\bm{v}_1,\bm{v}_2)$ is written as \cite{Takada2016}
\begin{equation}
\begin{cases}
	\bm{v}_1 = \bm{v}_1^\pp -A(\bm{v}_{12}^\pp \cdot \hat{\bm{k}})\hat{\bm{k}}\\
	\bm{v}_2 = \bm{v}_2^\pp +A(\bm{v}_{12}^\pp \cdot \hat{\bm{k}})\hat{\bm{k}}
\end{cases}\label{eq:v_change}
\end{equation}
where
\begin{equation}
A=
\begin{cases}
\displaystyle 1-\frac{1}{2}(1-e) \mathfrak{N}^2 \frac{\cos^2\theta_{\rm c}}{\cos^2\theta}& (\tilde{b}\le \min(\lambda,\mathfrak{N}))\\
1 & (\tilde{b} > \min(\lambda,\mathfrak{N}))
\end{cases},\label{eq:A}
\end{equation}
with the angle $\theta$ between $\bm{v}_{12}$ and $\hat{\bm{k}}$, and $\theta_{\rm c}$ satisfies $\cos\theta_{\rm c} = \{1-b^2/(\mathfrak{N}^2 d^2)\}^{1/2}$.
This process is equivalent to that used in Ref.\ \cite{Takada2016}.
It should be noted that the expression (\ref{eq:A}) is only valid for nearly elastic cases $1-e\ll 1$.

Let us consider a uniformly sheared flow characterized by $u_x=\dot\gamma y$, $u_y=u_z=0$ to derive observables in the steady state.
Using the peculiar velocity as $V_x=v_x-\dot\gamma y$, $V_y=v_y$, $V_z=v_z$, we can rewrite the Boltzmann equation (\ref{eq:Boltzmann}) as 
\begin{equation}
\left(\partial_t - \dot\gamma V_{1y} \frac{\partial}{\partial V_{1x}}\right) f(\bm{V}_1,t)
=J(\bm{V}_1|f),\label{eq:Boltzmann_shear}
\end{equation}
where we have ignored the spatial fluctuations in Eq.\ (\ref{eq:Boltzmann}).
Multiplying $mV_{1\alpha}V_{1\beta}$ with Eq.\ (\ref{eq:Boltzmann_shear}) and integrating over $\bm{V}_1$, we obtain the time evolution of the kinetic stress tensor
\begin{equation}
\partial_t P^k_{\alpha\beta} + \dot\gamma(\delta_{\alpha x}P^k_{y\beta} + \delta_{\beta x}P^k_{\alpha y})=-\Lambda_{\alpha\beta},\label{eq:P_Lambda}
\end{equation}
where $P^k_{\alpha\beta} \equiv \int d\bm{v} m V_{\alpha} V_{\beta} f(\bm{V},t)$ is the kinetic stress tensor and $\Lambda_{\alpha\beta}$ is defined by
\begin{equation}
\overleftrightarrow{\Lambda}\equiv -m \int d\bm{v}_1 \bm{V}_1 \bm{V}_1 J(\bm{V}_1|f).\label{eq:Lambda_def}
\end{equation}

We assume that the velocity distribution function is given by Grad's moment method \cite{Grad1949,Herdegen1982,Jenkins1985a,Jenkins1985b,Tij2001,Garzo2002,Garzo2005,Kremer2011,Garzo2013,Chamorro2015,Hayakawa2016_1,Hayakawa2016}
\begin{equation}
f(\bm{V})=f_{\rm M}(\bm{V}) \left[1+\frac{m}{2T} \left(\frac{P^k_{\alpha\beta}}{p^k}-\delta_{\alpha\beta}\right)V_\alpha V_\beta\right],\label{eq:f_0th}
\end{equation}
where we adopt Einstein's rule for Greek indices where duplicated indices take summation over $x$, $y$, and $z$. 
Here, we have introduced the pressure $p^k$ defined by $p^k \equiv (P^k_{xx}+P^k_{yy}+P^k_{zz})/3$ which satisfies the equation of state for the ideal gas $p^k=nT$.
We have also introduced the Maxwellian distribution function $f_{\rm M}(\bm{V})$: 
\begin{equation}
f_{\rm M}(\bm{V})=n \left(\frac{m}{2\pi T}\right)^{3/2} \exp\left(-\frac{mV^2}{2T}\right).
\end{equation}
Using the ansatz Eq.\ (\ref{eq:f_0th}), $\overleftrightarrow{\Lambda}$ can be divided into two parts:\ diagonal and non-diagonal parts as
\begin{align}
\overleftrightarrow{\Lambda}
&=\nu (\overleftrightarrow{P^k}-p^k \overleftrightarrow{1}) + \zeta p^k \overleftrightarrow{1}\label{eq:Lambda_12}
\end{align}
(the derivation is given in Appendix \ref{sec:Lambda}),
where $1_{\alpha\beta}=1$ for $\alpha=\beta$ and $0$ otherwise. 
Here, $\nu$ and $\zeta$ are, respectively, the frequency given by Eq.\ (\ref{eq:nu}) and the dissipation rate given by Eq.\ (\ref{eq:zeta}) in Appendix \ref{sec:zeta_nu}.

%%%%%%%%%%%%%%%%%%%%%%%%%%%%%%
\begin{figure}[htbp]
	\begin{center}
		\includegraphics[width=80mm]{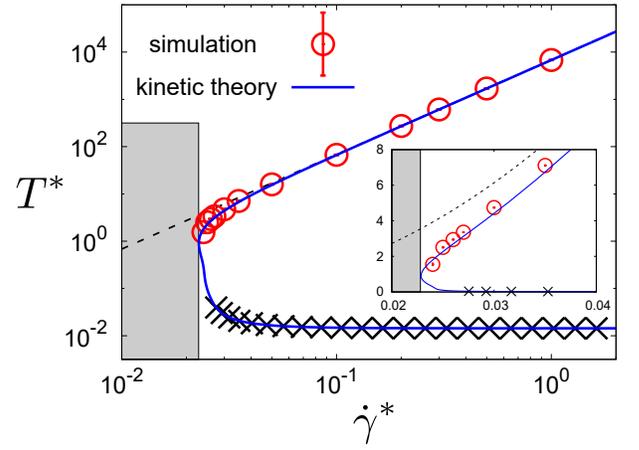}
	\end{center}
	\caption{The relationship between the temperature and the shear rate (open circles) and that from the kinetic theory (solid line) for $e=0.99$ and $\lambda=1.5$.
	The dashed lines expresses Bagnoldian scaling (\ref{eq:T_Bagnold}).
	The crosses show the linearly unstable steady solution.
	The shaded area exhibits an unreachable region, where there is no steady state.
	The inset shows the zoom in the vicinity of the critical shear rate.}
	\label{fig:temp}
\end{figure}
%%%%%%%%%%%%%%%%%%%%%%%%%%%%%%
%%%%%%%%%%%%%%%%%%%%%%%%%%%%%%
\begin{figure}[htbp]
	\begin{center}
		\includegraphics[width=80mm]{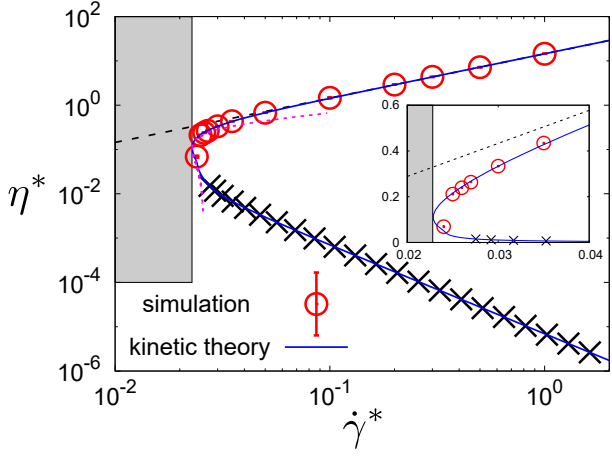}
	\end{center}
	\caption{The shear rate dependence of the shear viscosity for $e=0.99$ and $\lambda=1.5$.
	The dashed lines represents Bagnoldian scaling (\ref{eq:eta_Bagnold}).
	The cross points express the linearly unstable region.
	The dotted line shows the expansion around the critical shear rate $\dot\gamma_{\rm c}$ given by the quadratic function (\ref{eq:eta_c_gamma_c}).
	The shaded area exhibits an unreachable region, where there is no steady state.
	The inset shows the zoom in the vicinity of the critical shear rate.}
	\label{fig:steady_gamma_eta}
\end{figure}
%%%%%%%%%%%%%%%%%%%%%%%%%%%%%%

From Eqs.\ (\ref{eq:P_Lambda}) and (\ref{eq:Lambda_12}), we obtain the time evolution equation of the pressure, the normal stress difference $\Delta P^k\equiv P^k_{xx}-P^k_{yy}$, and the shear stress $P^k_{xy}$ as
\begin{align}
&\partial_t p^k+\frac{2}{3}\dot\gamma P^k_{xy}
	=-\zeta p^k,\label{eq:time_evol_p}\\
& \partial_t \Delta P^k+2\dot\gamma P^k_{xy} =-\nu \Delta P^k,\label{eq:time_evol_delp}\\
& \partial_t P^k_{xy}+\dot\gamma \left(p^k-\frac{1}{3}\Delta P^k\right) =-\nu P^k_{xy}.\label{eq:time_evol_P}
\end{align}
Unfortunately, the observables in Eqs.\ (\ref{eq:time_evol_p})--(\ref{eq:time_evol_P}) cannot be expressed as functions of $\dot\gamma$ explicitly.
We note that the second normal difference $P^k_{yy}-P^k_{zz}$ is not included in the above treatment, which is known to exist not only in denser systems \cite{Sangani1996,Hayakawa2017} but also in dilute systems \cite{Tsao1995}.
However, as shown in Appendix \ref{sec:N1_N2}, $P^k_{yy}-P^k_{zz}$ is much smaller than $\Delta P^k$ in this system.
Moreover, if we adopt the linearized approximation from the Maxwellian in the evaluation of the nonlinear collision integral, the second normal stress difference disappears in the dilute gas.
This is the reason why we only consider the set of $p^k$, $P^k_{xy}$, and $\Delta P^k$.

%%%%%%%%%%%%%%%%%%%%%%%%%%%%%%
\begin{figure}[htbp]
	\begin{center}
		\includegraphics[width=80mm]{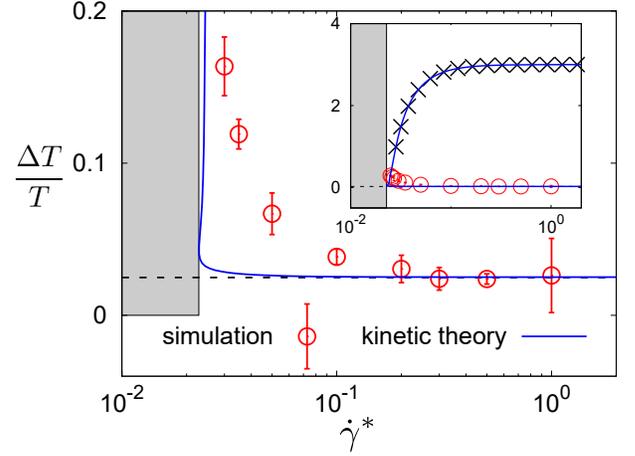}
	\end{center}
	\caption{The relationship between the temperature difference and the shear rate obtained from the simulation (red open circles) and that from the kinetic theory (blue solid line) for $e=0.99$ and $\lambda=1.5$.
	The black dashed lines exhibits Bagnoldian scaling.
	The shaded area exhibits an unreachable region, where there is no steady state.
	The inset shows the zoom in the vicinity of the critical shear rate.}
	\label{fig:del_T}
\end{figure}
%%%%%%%%%%%%%%%%%%%%%%%%%%%%%%

%%%%%%%%%%%%%%%%%%%%%%%%%%%%%%
\begin{figure}[htbp]
	\begin{center}
		\includegraphics[width=80mm]{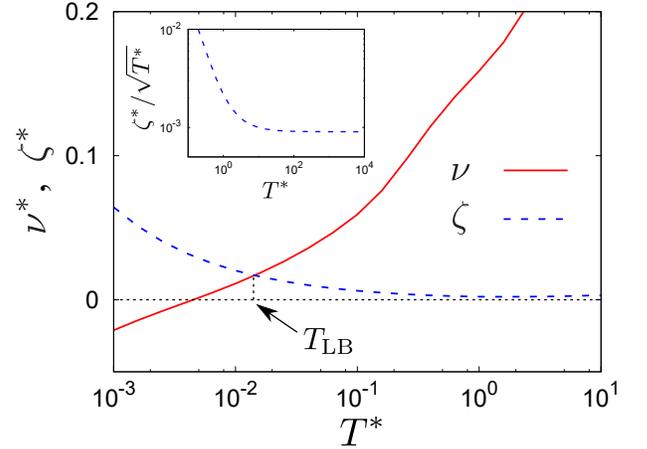}
	\end{center}
	\caption{The temperature dependence of $\nu$ (solid line) and $\zeta$ (dashed line) for $e=0.99$ and $\lambda=1.5$.
	The inset shows the temperature dependence of $\zeta/\sqrt{T}$ for high temperature.}
	\label{fig:nu}
\end{figure}
%%%%%%%%%%%%%%%%%%%%%%%%%%%%%%

Let us focus on the steady state.
From Eqs.\ (\ref{eq:time_evol_p})--(\ref{eq:time_evol_P}), we can express the shear rate, shear stress tensor, and stress difference as a function of $T$ in the steady state as
\begin{align}
	&\dot\gamma=\sqrt{\frac{3}{2}\frac{\nu^2\zeta}{\nu-\zeta}}\nonumber\\
	&=nd^2\sqrt{\frac{\pi \varepsilon}{m}}\left(\frac{\varepsilon}{T}\right)^3
	\left[(1-e)^{1/2}\dot\gamma^{(1)*}+\mathcal{O}\left((1-e)\right)\right],\label{eq:steady_gamma}\\
	%%%
	&P^k_{xy}=-\frac{p^k}{\nu}\sqrt{\frac{3}{2}\zeta(\nu-\zeta)}\nonumber\\
	&=n\varepsilon \left(\frac{T}{\varepsilon}\right)^{3/2}
	\left[(1-e)^{1/2}P_{xy}^{k(1)*}+\mathcal{O}\left((1-e)\right)\right],\label{eq:steady_Pxy}
	%%%
\end{align}
\begin{align}
	&\Delta P^k=\frac{3\zeta}{\nu}p^k\nonumber\\
	&=3n\varepsilon \left(\frac{T}{\varepsilon}\right)^{3/2}
	\left[(1-e)\Delta P^{*}(T)+\mathcal{O}\left((1-e)^2\right)\right],
\end{align}
respectively, where the quantities with asterisk are dimensionless variables such as
\begin{align}
	\dot\gamma^{(1)*}&\equiv \sqrt{\frac{3}{2}\nu^{(0)*}_1 \zeta^{(1)*}_1},\\
	P_{xy}^{k(1)*}&\equiv \frac{1}{{\nu^{(0)*}_1}}\sqrt{\frac{3}{2} \nu^{(0)*}_1(T)\zeta^{(1)*}_1},\\
	\Delta P^{(1)*}&\equiv \frac{\zeta^{(1)*}_1}{\nu^{(0)*}_1}.
\end{align}
Here, $\nu_1^{(0)*}$ and $\zeta_1^{(1)*}$ are given by Eqs.\ (\ref{eq:nu_0_1}) and (\ref{eq:zeta_1_1}), respectively.
From Eqs.\ (\ref{eq:steady_gamma}) and (\ref{eq:steady_Pxy}), we obtain the shear viscosity
\begin{align}
&\eta =\frac{(\nu-\zeta)p^k}{\nu^2}\nonumber\\
	&= \frac{1}{d^2}\sqrt{\frac{m\varepsilon}{\pi}} \left(\frac{T}{\varepsilon}\right)^{9/2}\nonumber\\
	&\hspace{1em}\times \left[\eta^{(0)*}+(1-e)\eta^{(1)*}+\mathcal{O}\left((1-e)^2\right)\right],\label{eq:steady_eta}
\end{align}
where
\begin{align}
\eta^{(0)*}&\equiv \frac{1}{\nu^{(0)*}_1},\\
\eta^{(1)*}&\equiv -\frac{1}{\nu^{(0)*2}_1}\left[\nu^{(1)*}_1+\nu^{(1)*}_2 +\frac{T}{\varepsilon}\zeta^{(1)*}_1\right].\label{eq:eta_1}
\end{align}
Here, $\nu_1^{(1)*}$ and $\nu_2^{(1)*}$ are given by Eqs.\ (\ref{eq:nu_1_1}) and (\ref{eq:nu_1_2}), respectively.
We note that Eq.\ (\ref{eq:steady_gamma}) determines the relationship between the shear rate and the temperature, where it is easy to express the shear rate as a function of the temperature, though the actual control parameter is the shear rate.
Then, the relationships between the shear rate and the other observables are also parametrically plotted in terms of the temperature in Figs.\ \ref{fig:temp}--\ref{fig:del_T}.
Figure \ref{fig:nu} plots the temperature dependence of $\nu^*\equiv \nu\sqrt{md^2/\varepsilon}$ and $\zeta^*\equiv \zeta\sqrt{md^2/\varepsilon}$ whose expressions are presented in Appendix \ref{sec:zeta_nu}.
The steady expressions in the simple shear flow can only exist above the lower bound temperature $T_{\rm LB}$ which is determined by $\nu(T)=\zeta(T)$ (see Eqs.\ (\ref{eq:steady_gamma}) and (\ref{eq:steady_Pxy})).
For our choice of parameters $e=0.99, \lambda=1.5$, the lower bound temperature is given by $T_{\rm LB}^*(\equiv T_{\rm LB}/\varepsilon)=0.0144$.

There are two branches for the theoretical $\eta$ above the critical shear rate, though $\eta$ on the lower branch is linearly unstable as explained in Appendix \ref{sec:stability}.
The shear rate and the stable viscosity, respectively, tend to $\dot\gamma^2/\omega_{\rm HC}^2 \to (5/4)(1-e)$ and $\eta/\eta_{\rm HC}\to 1-(5/6)(1-e)$ for $T>T_{{\rm c}}$ and $e\to 1$, where $\omega_{\rm HC}=(16/5)nd^2\sqrt{\pi T/m}$ and $\eta_{\rm HC}=5/(16d^2)\sqrt{mT/\pi}$ are the collision frequency and the shear viscosity of dilute hard-core gases, respectively \cite{Santos2004}.
It is remarkable that the upper branches in Figs.\ \ref{fig:temp} and \ref{fig:steady_gamma_eta} reproduces well the simulation results.
We also note that the temperature difference $\Delta T=\Delta P^k/n$ obtained from both the simulation and the kinetic theory of hard-core dilute granular gases
\begin{equation}
	\Delta T=\frac{25\pi(2+e)}{432(1+e)^2(3-e)^3}\frac{1}{\varphi^2}md^2\dot\gamma^2 \label{eq:del_T_Bagnold}
\end{equation}
agree in high sheared regime, though the theoretical prediction does not agree with the simulation in the low shear regime (this relationship can be derived easily by Ref.\ \cite{Santos2004}), while the deviations become large near the critical shear rate $\dot\gamma \approx \dot\gamma_{\rm c}$.
Then, $\Delta T/T=5(1-e)/(3-e)$ with the aid of Eq.\ (\ref{eq:T_Bagnold}) (Fig.\ \ref{fig:del_T}).

Let us evaluate the critical shear viscosity $\dot\gamma_{\rm c}$ below which the steady state does not exist.
The critical condition is given by $\partial \eta /\partial \dot\gamma = (\partial \eta /\partial T) / (\partial \dot\gamma /\partial T) \to \infty$ at $\dot\gamma_{\rm c}$, which is reduced to $\partial \dot\gamma /\partial T=0$. 
As shown in Appendix \ref{sec:critical_temp}, this critical temperature $T_{\rm c}(\equiv T_{\rm c}/\varepsilon)=0.910$ corresponds to the critical shear rate  $\dot\gamma_{\rm c}^*=0.0228$.
The theoretical critical shear viscosity agrees with the numerical critical shear viscosity.

We perform the linear stability analysis for the sheared uniform state (the detailed explanation is given in Appendix \ref{sec:stability}).
For simplicity, we ignore the spatial degree of freedom.
From the set of equations (\ref{eq:time_evol_p})--(\ref{eq:time_evol_P}), the uniform shear state is stable for $T>T_{\rm CL}(=0.04\varepsilon)$ in our choice of parameters.
We plot the linearly unstable region as crosses in Figs.\ \ref{fig:temp}--\ref{fig:del_T}.

%%%%%%%%%%%%%%%%%%%%%%%%

\section{Conclusion}\label{sec:summary}
In this paper, we have performed the event-driven molecular dynamics simulation for cohesive granular gases under a uniform shear and clarified the rheological properties of these particles.
We have found that there exists a steady state when the shear rate is larger than the critical shear rate, while the clustering proceeds for lower shear rate.
Even for lower shear rate, introducing the effective shear rate and the shear viscosity, we have found that the flow curve can be approximately expressed as the Bagnoldian expression if we replace the diameter of the particle by the mean diameter of clusters. 
We have obtained two branches for the steady uniformly sheared state from the analysis of the inelastic Boltzmann equation, one of which is consistent with the simulation, and the other branch is linearly unstable.

\section*{Acknowledgments}
The authors thank Andr\'{e}s Santos, Koshiro Suzuki, Kuniyasu Saitoh, Takeshi Kawasaki, and Michio Otsuki for their useful comments.
One of the authors (ST) wishes to express his sincere gratitude to Tomohiko G.\ Sano and Thorsten P\"{o}schel for their helpful comments.
Numerical computation in this work was partially carried out at the Yukawa Institute Computer Facility.
This work is partially supported by Scientific Grant-in-Aid of JSPS, KAKENHI (Grant No.\ 16H04025 and No.\ JP16H06478).

%%%%%%%%%%%%%%%%%%%%%%%%%%%%%%
%%%%%%%%%%%%%%%%%%%%%%%%%%%%%%
\appendix
\section{Simulation under the flat boundary condition}\label{sec:LEbc}
%%%%%%%%%%%%%%%%%%%%%%%%%%%%%%
\begin{figure}[htbp]
	\begin{center}
		\includegraphics[width=80mm]{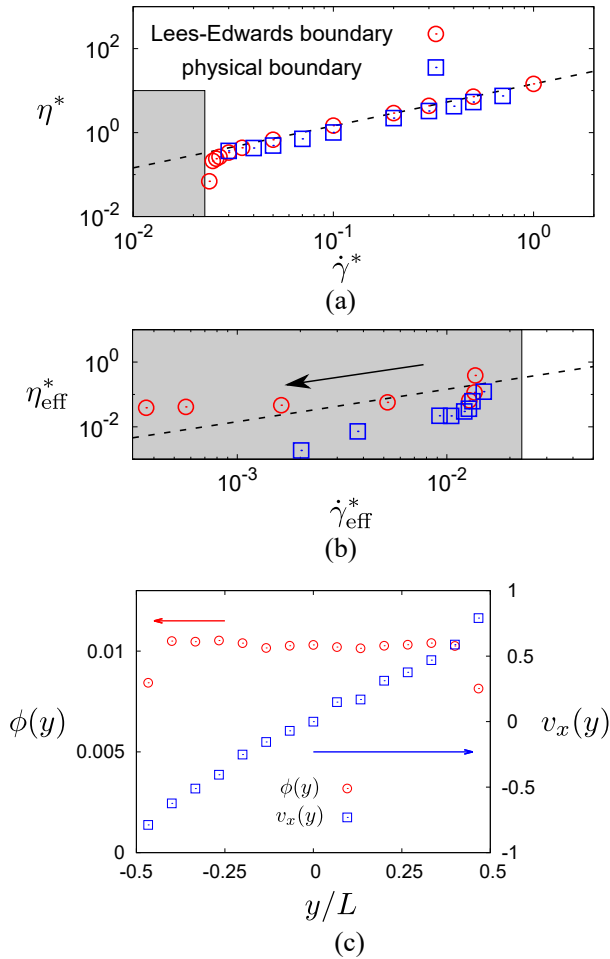}
	\end{center}
	\caption{(a) We present the shear viscosity against the shear rate obtained from the simulation under the Lees-Edwards boundary condition (open circles) and those under the flat boundary condition (open squares) for $e=0.99$ and $\lambda=1.5$.
	(b) We also plot the effective shear viscosity Eq.\ (\ref{eq:eff_vis_cl}) against the effective shear rate Eq.\ (\ref{eq:gamma_vis_cl}) for $\dot\gamma^*=0.015$.
	The dashed line expresses Bagnoldian expressions Eq.\ (\ref{eq:eta_Bagnold}).
	The shaded area expresses the region which does not have any steady state.
	The arrow indicates the time evolution.
	(c) We plot the density and velocity profiles obtained from the simulation under the physical boundary condition for $\dot\gamma^*=0.05$.}
	\label{fig:physical_boundary}
\end{figure}
%%%%%%%%%%%%%%%%%%%%%%%%%%%%%%
In this Appendix, we examine the applicability of the Lees-Edwards boundary condition from the comparison of the simulations under the flat boundary condition \cite{Takada2014}.
We prepare two flat walls at $y=\pm L/2$, moving in $x$-direction with the velocities $\pm \dot\gamma L/2$.
When a particle having $(v_x,v_y,v_z)$ hits the walls at $y=\pm L/2$, the velocity changes to $(v_x \pm \dot\gamma L/2, -v_y,v_z)$ after the collision, respectively.

Figure \ref{fig:physical_boundary} plots (a) the shear viscosity against the shear rate (\ref{eq:gamma_vis_cl}) for $\dot\gamma > \dot\gamma_{\rm c}$ and (b) the effective flow curve for $\dot\gamma < \dot\gamma_{\rm c}$ which gives the relationship between $\eta_{\rm eff}$ and $\dot\gamma_{\rm eff}$.
The result under the flat boundary condition for $\dot\gamma>\dot\gamma_{\rm c}$ almost agrees with that under the Lees-Edwards boundary condition and the Bagnoldian expression (\ref{eq:eta_Bagnold}), which is contrary to the previous studies under the bumpy boundary condition \cite{Liem1992, Santos1992, Tij2001}.
For $\dot\gamma > \dot\gamma_{\rm c}$, the density and velocity gradient are almost constant except for the boundary layers as shown in Fig.\ \ref{fig:physical_boundary}(c).
We also note that the effective viscosity obtained from the simulation under the flat boundary condition becomes much smaller than that under the Lees-Edwards boundary condition for $\dot\gamma<\dot\gamma_{\rm c}$.
In conclusion, to adopt the Lees-Edwards boundary condition does not cause any artifact for $\dot\gamma>\dot\gamma_{\rm c}$ while it is controversial for $\dot\gamma<\dot\gamma_{\rm c}$.

%%%%%%%%%%%%%%%%%%%%%%%%%%%%%%

%%%%%%%%%%%%%%%%%%%%%%%%%%%%%%
\section{Results for highly dissipative cases}\label{sec:e_change}
%%%%%%%%%%%%%%%%%%%%%%%%%%%%%%
\begin{figure}[htbp]
	\begin{center}
		\includegraphics[width=80mm]{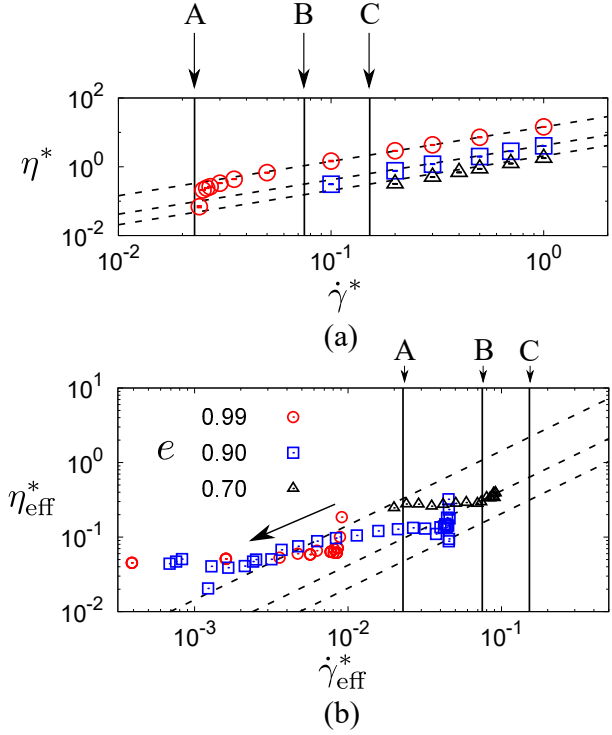}
	\end{center}
	\caption{(a) The relationships between the shear viscosity and the shear rate for various restitution coefficient $e=0.99$ (open circles), $0.90$ (open squares), and $0.70$ (open triangles) and $\lambda=1.5$.
	The dashed lines express Bagnoldian expressions Eq.\ (\ref{eq:eta_Bagnold}) for various $e$.
	(b) We present the effective shear viscosity Eq.\ (\ref{eq:eff_vis_cl}) against the effective shear rates Eq.\ (\ref{eq:gamma_vis_cl}) for $(e, \dot\gamma^*)=(0.99, 0.01)$ (filled circles), $(0.90, 0.05)$ (filled squares), and $(0.70, 0.1)$ (filled triangles).
	The arrow indicates the time evolution.
	The vertical lines indicated by A, B, and C shows the critical shear rate $\dot\gamma^*_{\rm c}(e)$ for various $e=0.99$, $0.90$, and $0.70$, respectively.
	Here, $\dot\gamma^*_{\rm c}(e=0.99)=0.0228$, $\dot\gamma^*_{\rm c}(e=0.90)=0.0745$, and $\dot\gamma^*_{\rm c}(e=0.70)=0.152$.}
	\label{fig:eta_e_change}
\end{figure}
%%%%%%%%%%%%%%%%%%%%%%%%%%%%%%
In this Appendix, we present the results for larger inelasticity, especially for $e=0.9$, and $0.7$.
Figure \ref{fig:eta_e_change} shows the results of the shear viscosity against the shear rate and the effective shear viscosity (\ref{eq:eff_vis_cl}) against the effective shear rate (\ref{eq:gamma_vis_cl}).
Even for moderately inelastic case, Bagnoldian expressions (\ref{eq:T_Bagnold}) and (\ref{eq:eta_Bagnold}) give precise results if the steady state exists above the critical shear rate.

However, the time evolutions of the temperature and the effective shear viscosity (\ref{eq:eff_vis_cl}) against the effective shear rate (\ref{eq:gamma_vis_cl}) in Fig.\ \ref{fig:eta_e_change} cannot be used, if the steady state is unstable for $\dot\gamma< \dot\gamma_{\rm c}$ for larger inelasticity.

%%%%%%%%%%%%%%%%%%%%%%%%%%%%%%
\section{Comparison of the stress tensor with the kinetic stress tensor}\label{sec:ratio_Pk_P}
%%%%%%%%%%%%%%%%%%%%%%%%%%%%%%
\begin{figure}[htbp]
	\begin{center}
		\includegraphics[width=80mm]{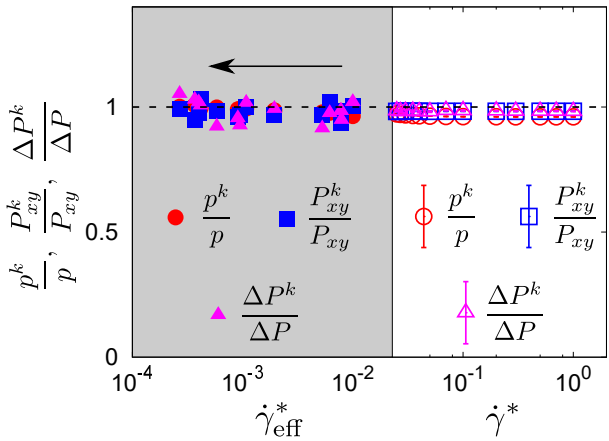}
	\end{center}
	\caption{The ratios of the pressure, the shear stress, and the pressure difference to the kinetic part of each of them for $e=0.99$.
	The arrow indicates the time evolution.}
	\label{fig:P_ratio}
\end{figure}
%%%%%%%%%%%%%%%%%%%%%%%%%%%%%%
In this Appendix, we verify whether the stress tensor is dominated by the kinetic stress.
To validate this, we measure the ratio of the stress tensor (\ref{eq:P_measurement}) to the kinetic part of the stress tensor by using our full scratched simulation code.
Here, the stress tensor is defined as
\begin{equation}
	\overleftrightarrow{P}=\overleftrightarrow{P}^k + \overleftrightarrow{P}^c,\label{eq:def_P}
\end{equation}
where $\overleftrightarrow{P}^c$ is the collisional contribution of the stress tensor, defined by
\begin{equation}
	\overleftrightarrow{P}^c=\frac{1}{L^3\Delta \tau}\sum_{i,j}^{\rm event}\Delta \bm{p}_{ij} \bm{r}_{ij}.\label{eq:def_Pc}
\end{equation}
Here, $\Delta \bm{p}_i$ is the change of the momentum of $i$-th particle during a collision, and $\Delta \tau$ is set to $100\sqrt{md^2/\varepsilon}$.
We have confirmed that the results are insensitive to the time interval $\Delta \tau$ in the range $10\sqrt{md^2/\varepsilon}$ to $100\sqrt{md^2/\varepsilon}$.
Figure \ref{fig:P_ratio} represents the ratios for the pressure, the shear stress, and the pressure difference.
All quantities satisfy
\begin{equation}
	0.98< \frac{P^k}{P},\ \frac{P^k_{xy}}{P_{xy}},\ \frac{\Delta P^k}{\Delta P} \le 1,
\end{equation}
for $\dot\gamma>\dot\gamma_{\rm c}$, and
\begin{equation}
	0.9< \frac{P^k}{P},\ \frac{P^k_{xy}}{P_{xy}},\ \frac{\Delta P^k}{\Delta P} \le 1.1,
\end{equation}
for $\dot\gamma<\dot\gamma_{\rm c}$.
in our simulations.
Although the data are a little scattered for the unstable regime ($\dot\gamma<\dot\gamma_{\rm c}$), we conclude that the collisional contribution to the stress tensor is negligible even for $\dot\gamma < \dot\gamma_{\rm c}$.

%%%%%%%%%%%%%%%%%%%%%%%%%%%%%%
\section{Derivation of Eq.\ (\ref{eq:Lambda_12})}\label{sec:Lambda}
In this Appendix, let us show the detailed derivation of $\overleftrightarrow{\Lambda}$.
From the collision rule Eq.\ (\ref{eq:v_change}), the following relationship is satisfied:
\begin{align}
&\bm{c}_1^\prime\bm{c}_1^\prime + \bm{c}_2^\prime\bm{c}_2^\prime - \bm{c}_1\bm{c}_1 + \bm{c}_2\bm{c}_2\nonumber\\
&=-A(\bm{c}_{12}\cdot \hat{\bm{k}})(\bm{c}_{12} \hat{\bm{k}}+ \hat{\bm{k}}\bm{c}_{12})
	+2A^2 (\bm{c}_{12}\cdot \hat{\bm{k}})^2\hat{\bm{k}}\hat{\bm{k}}.\label{eq:c1c2_change}
\end{align}
Here, we introduce the dimensionless velocity $\bm{c}_i \equiv \bm{v}_i / (2\varepsilon/m)^{1/2}$.
Inserting Eq.\ (\ref{eq:c1c2_change}) into Eq.\ (\ref{eq:Lambda_def}) and integrating over $\bm{C}=(\bm{c}_1+\bm{c}_2)/2$, $\Lambda$ is expressed as
\begin{align}
\overleftrightarrow{\Lambda}
&= -\frac{1}{2}\pi^{-3}mn^2d^2 \left(\frac{\pi \varepsilon}{m}\right)^{3/2} \left(\frac{\varepsilon}{T}\right)^{5/2}\nonumber\\
&\hspace{1em}\times \int d\bm{c}_{12}\int d\hat{\bm{k}} \tilde{\sigma}(\chi,c_{12})c_{12} \exp\left(-\frac{1}{2}c_{12}^2\right)\nonumber\\
&\hspace{1em}\times \left[1+\frac{1}{2}(P_{\alpha\beta}^*-\delta_{\alpha\beta}) c_{12,\alpha}c_{12,\beta} \right]\nonumber\\
&\hspace{1em}\times \left[-A(\bm{c}_{12}\cdot \hat{\bm{k}})(\bm{c}_{12} \hat{\bm{k}}+ \hat{\bm{k}}\bm{c}_{12})
	+2A^2 (\bm{c}_{12}\cdot \hat{\bm{k}})^2\hat{\bm{k}}\hat{\bm{k}}\right]\nonumber\\
&\equiv -\frac{1}{2}mn^2d^2 \left(\frac{\varepsilon}{\pi m}\right)^{3/2} \left(\frac{\varepsilon}{T}\right)^{5/2}
	\left(\overleftrightarrow{\Lambda_1^*} + \overleftrightarrow{\Lambda_2^*} 
	+ \overleftrightarrow{\Lambda_3^*} + \overleftrightarrow{\Lambda_4^*}\right),\label{eq:Lambda_sum}
\end{align}
where $\bm{c}_{12}=\bm{c}_1-\bm{c}_2$, $P^*_{\alpha\beta}=P_{\alpha\beta}/(nT)$, and we have introduced $\overleftrightarrow{\Lambda_i^*}$ ($i=1,2,3,4$) as
\begin{align}
\overleftrightarrow{\Lambda_1^*}
&= -\int d\bm{c}_{12}\int d\hat{\bm{k}} \tilde{\sigma}(\chi,c_{12})c_{12} \exp\left(-\frac{\varepsilon}{2T}c_{12}^2\right)\nonumber\\
&\hspace{1em}\times A(\bm{c}_{12}\cdot \hat{\bm{k}})(\bm{c}_{12} \hat{\bm{k}}+ \hat{\bm{k}}\bm{c}_{12}),\label{eq:Lambda1}\\
%%%%%%%%%%%%%%%%%%%%%%%%%
\overleftrightarrow{\Lambda_2^*}
&= 2\int d\bm{c}_{12}\int d\hat{\bm{k}} \tilde{\sigma}(\chi,c_{12})c_{12} \exp\left(-\frac{\varepsilon}{2T}c_{12}^2\right)\nonumber\\
&\hspace{1em}\times A^2 (\bm{c}_{12}\cdot \hat{\bm{k}})^2\hat{\bm{k}}\hat{\bm{k}},\label{eq:Lambda2}\\
%%%%%%%%%%%%%%%%%%%%%%%%%
\overleftrightarrow{\Lambda_3^*}
&= -\frac{1}{2} \int d\bm{c}_{12}\int d\hat{\bm{k}} \tilde{\sigma}(\chi,c_{12})c_{12} \exp\left(-\frac{\varepsilon}{2T}c_{12}^2\right)\nonumber\\
&\hspace{1em}\times (P_{\alpha\beta}^*-\delta_{\alpha\beta})c_{12,\alpha}c_{12,\beta} A(\bm{c}_{12}\cdot \hat{\bm{k}})(\bm{c}_{12} \hat{\bm{k}}+ \hat{\bm{k}}\bm{c}_{12}),\label{eq:Lambda3}\\
%%%%%%%%%%%%%%%%%%%%%%%%%
\overleftrightarrow{\Lambda_4^*}
&= \int d\bm{c}_{12}\int d\hat{\bm{k}} \tilde{\sigma}(\chi,c_{12})c_{12} \exp\left(-\frac{\varepsilon}{2T}c_{12}^2\right)\nonumber\\
&\hspace{1em}\times (P_{\alpha\beta}^*-\delta_{\alpha\beta})c_{12,\alpha}c_{12,\beta} A^2 (\bm{c}_{12}\cdot \hat{\bm{k}})^2\hat{\bm{k}}\hat{\bm{k}},\label{eq:Lambda4}
\end{align}
respectively.
Here, $\chi$ is the scattering angle and we expand $\chi$ in terms of the small inelasticity $1-e$ as
\begin{equation}
	\chi=\chi^{(0)}+(1-e) \chi^{(1)}+\mathcal{O}\left((1-e)^2\right).
\end{equation}
The explicit forms of $\chi^{(0)}$ and $\chi^{(1)}$ are, respectively, given by \cite{Takada2016}
\begin{align}
\chi^{(0)}&=
\begin{cases}
	\chi_{\rm inelastic}^{(0)} & (\tilde{b} \le \min(\lambda,\mathfrak{N}))\\
	\chi_{\rm grazing}^{(0)} & (\min(\lambda,\mathfrak{N})<\tilde{b}\le \lambda)\\
	0 & (\tilde{b}>\lambda)
\end{cases},\\
\chi^{(1)}&=
\begin{cases}
	\chi_{\rm inelastic}^{(1)} & (\tilde{b} \le \min(\lambda,\mathfrak{N}))\\
	0 & (\tilde{b} > \min(\lambda,\mathfrak{N}))
\end{cases},
\end{align}
with the dimensionless collision parameter $\tilde{b}\equiv b/d$ and
\begin{align}
	\chi_{\rm inelastic}^{(0)} &=\pi - 2\sin^{-1}\frac{\tilde{b}}{\lambda} -2\sin^{-1}\frac{\tilde{b}}{\mathfrak{N}} +2\sin^{-1}\frac{\tilde{b}}{\mathfrak{N}\lambda},\\
	\chi_{\rm grazing}^{(0)} &= 2\sin^{-1}\frac{\tilde{b}}{\mathfrak{N} \lambda} - 2\sin^{-1} \frac{\tilde{b}}{\lambda},\\
	\chi_{\rm inelastic}^{(1)} &= \displaystyle -\left[\frac{\tilde{b}\mathfrak{N}^2}{\sqrt{\lambda^2 -\tilde{b}^2}}
		+ \frac{\tilde{b}}{\sqrt{\mathfrak{N}^2 -\tilde{b}^2}}
		- \frac{\tilde{b}}{\sqrt{\mathfrak{N}^2\lambda^2 -\tilde{b}^2}}\right]\nonumber\\
		&\hspace{1em}\times\cos^2\theta_{\rm c}.\label{eq:chi1}
\end{align}

To evaluate $\overleftrightarrow{\Lambda_i^*}$ ($i=1,2,3,4$), the following relations are useful:
\begin{align}
&\int d\hat{\bm{k}} \tilde{\sigma}(\chi,c_{12}) (\bm{c}_{12}\cdot \hat{\bm{k}})\hat{\bm{k}} 
= 2\pi \int_0^\infty d\tilde{b} \hspace{0.2em}\tilde{b} \sin^2\frac{\chi}{2} \bm{c}_{12},\label{eq:k1}\\
&\int d\hat{\bm{k}} \tilde{\sigma}(\chi,c_{12}) (\bm{c}_{12}\cdot \hat{\bm{k}})^2\hat{\bm{k}}\hat{\bm{k}} \nonumber\\
&=\pi \int_0^\infty d\tilde{b} \hspace{0.2em}\tilde{b} \sin^2\frac{\chi}{2} \nonumber\\
	&\hspace{1em}\times\left[c_{12}^2 \cos^2\frac{\chi}{2}\overleftrightarrow{1}+\left(2\sin^2\frac{\chi}{2}-\cos^2\frac{\chi}{2}\right)\bm{c}_{12}\bm{c}_{12}\right].\label{eq:k2}
\end{align}
It should be noted that these relations can be derived if we choose $\bm{c}_{12}$ as $z$-axis and use polar coordinates to $\hat{\bm{k}}$.
Inserting Eq.\ (\ref{eq:k1}) into Eq.\ (\ref{eq:Lambda1}), we rewrite $\overleftrightarrow{\Lambda_1^*}$ as
\begin{align}
\overleftrightarrow{\Lambda_1^*}
&= -4\pi \int d\bm{c}_{12}\int_0^\infty d\tilde{b} \nonumber\\
	&\hspace{4em}\times A\tilde{b} c_{12} \sin^2\frac{\chi}{2}\exp\left(-\frac{\varepsilon}{2T}c_{12}^2\right) \bm{c}_{12}\bm{c}_{12}\nonumber\\
&= -\frac{16}{3}\pi^2 \int_0^\infty dc_{12}\int_0^\infty d\tilde{b} \nonumber\\
	&\hspace{5em}\times A\tilde{b} c_{12}^5
	\sin^2\frac{\chi}{2}\exp\left(-\frac{\varepsilon}{2T}c_{12}^2\right) \overleftrightarrow{1}.\label{eq:Lambda1_}
\end{align}
Similarly, from Eqs.\ (\ref{eq:Lambda2}) and (\ref{eq:k2}), $\overleftrightarrow{\Lambda_2^*}$ reduces to
\begin{align}
\overleftrightarrow{\Lambda_2^*}
&= 2\pi \int d\bm{c}_{12}\int_0^\infty d\tilde{b} A^2\tilde{b}c_{12} \sin^2\frac{\chi}{2}\exp\left(-\frac{\varepsilon}{2T}c_{12}^2\right)\nonumber\\
	&\hspace{1em}\times\left[c_{12}^2 \cos^2\frac{\chi}{2}\bm{1}+\left(2\sin^2\frac{\chi}{2}-\cos^2\frac{\chi}{2}\right)\bm{c}_{12}\bm{c}_{12}\right]\nonumber\\
&= \frac{16}{3}\pi^2 \int dc_{12}\int_0^\infty d\tilde{b} A^2\tilde{b}c_{12}^5 \sin^2\frac{\chi}{2}\exp\left(-\frac{\varepsilon}{2T}c_{12}^2\right)\overleftrightarrow{1}.
\label{eq:Lambda2_}
\end{align}
Similarly, $\overleftrightarrow{\Lambda_3^*}$ and $\overleftrightarrow{\Lambda_4^*}$ are, respectively, given by
\begin{align}
\overleftrightarrow{\Lambda_3^*}
&= -\frac{16}{15}\pi^2 \int dc_{12}\int_0^\infty d\tilde{b} \nonumber\\
	&\hspace{1em}\times A\tilde{b}c_{12}^7 \sin^2\frac{\chi}{2}\exp\left(-\frac{\varepsilon}{2T}c_{12}^2\right)\left(\overleftrightarrow{P}^*- \overleftrightarrow{1}\right),\label{eq:Lambda3_}\\
%%%%%%%%%%%%%%%%%%%%%%%%%
\overleftrightarrow{\Lambda_4^*}
&= \frac{16}{15}\pi^2 \int dc_{12}\int_0^\infty d\tilde{b} A\tilde{b}c_{12}^7 \sin^2\frac{\chi}{2}\nonumber\\
	&\hspace{1em}\times \left(1-\frac{3}{2}\cos^2\frac{\chi}{2}\right)\exp\left(-\frac{\varepsilon}{2T}c_{12}^2\right)\left(\overleftrightarrow{P}^*- \overleftrightarrow{1}\right).\label{eq:Lambda4_}
\end{align}
Inserting Eqs.\ (\ref{eq:Lambda1_})--(\ref{eq:Lambda4_}) into Eq.\ (\ref{eq:Lambda_sum}), we can obtain Eq.\ (\ref{eq:Lambda_12}).

%%%%%%%%%%%%%%%%%%%%%%%%%
\section{Derivation of the energy dissipation rate $\zeta$ and the frequency $\nu$}\label{sec:zeta_nu}
In this Appendix, we derive the expression for the energy dissipation rate $\zeta$ and the frequency $\nu$ in terms of the Boltzmann equation.

First, let us evaluate the energy dissipation rate $\zeta$.
The energy dissipation rate can be expanded in terms of the series of the small inelasticity $1-e$ as
\begin{align}
	\zeta(T)&=\frac{8}{3}nd^2 \sqrt{\frac{\pi \varepsilon}{m}} \left(\frac{\varepsilon}{T}\right)^{5/2}
	\int_0^\infty dc_{12} \int_0^\infty d\tilde{b}\nonumber\\
	&\hspace{1em}\times A(1-A)\tilde{b}c_{12}^5 \sin^2\frac{\chi}{2}\exp\left(-\frac{\varepsilon}{2T}c_{12}^2\right)\nonumber\\
	&\equiv \zeta^{(0)}(T)+(1-e) \zeta^{(1)}(T)+\mathcal{O}\left((1-e)^2\right),\label{eq:zeta}
\end{align}
where $\zeta^{(0)}$ and $\zeta^{(1)}$ are, respectively, given by \cite{Takada2016}
\begin{align}
	\zeta^{(0)}(T)&=0,\\
	\zeta^{(1)}(T)&=nd^2 \sqrt{\frac{\pi \varepsilon}{m}} \left(\frac{\varepsilon}{T}\right)^{5/2} \zeta_1^{(1)*}(T),
\end{align}
with
\begin{align}
\zeta_1^{(1)*}(T)&\equiv \frac{4}{3}\int_0^\infty dc_{12} \int_0^{\tilde{b}_{\rm max}} d\tilde{b}\nonumber\\
	&\hspace{1em}\times \tilde{b}\left(\mathfrak{N}^2-\tilde{b}^2\right)c_{12}^5 \exp\left(-\frac{\varepsilon}{2T}c_{12}^2\right),\label{eq:zeta_1_1}
\end{align}
where $\tilde{b}_{\rm max}=\min(\mathfrak{N},\lambda)$ with the introduction of a function $\min(x,y)$ to select the smaller one between $x$ and $y$.

Next, let us derive $\nu$.
From the Appendix \ref{sec:Lambda}, the frequency $\nu$ is written as
\begin{align}
	\nu (T)&= \frac{8}{15}nd^2\sqrt{\frac{\pi \varepsilon}{m}} \left(\frac{\varepsilon}{T}\right)^{7/2}
	\int_0^\infty dc_{12} \int_0^\infty d\tilde{b}A\tilde{b}c_{12}^7\nonumber\\
	&\hspace{1em}\times \left[(1-A)+\frac{3}{2}A\cos^2\frac{\chi}{2}\right] 
	\sin^2\frac{\chi}{2}\exp\left(-\frac{\varepsilon}{2T}c_{12}^2\right)\nonumber\\
	&\equiv \nu^{(0)}(T)+(1-e) \nu^{(1)}(T)+\mathcal{O}\left((1-e)^2\right),\label{eq:nu}
\end{align}
where $\nu^{(0)}$ and $\nu^{(1)}$ are, respectively, given by
\begin{align}
	\nu^{(0)}(T)&= nd^2\sqrt{\frac{\pi \varepsilon}{m}} \left(\frac{\varepsilon}{T}\right)^{7/2} \nu^{(1)*}_1(T),\\
	\nu^{(1)}(T)&= nd^2 \sqrt{\frac{\pi \varepsilon}{m}} \left(\frac{\varepsilon}{T}\right)^{7/2} 
	\left(\nu^{(1)*}_1(T)+\nu^{(1)*}_2(T)\right),
\end{align}
with
\begin{align}
\nu^{(0)*}_1(T)&= \frac{1}{5} 
	\int_0^\infty dc_{12} \int_0^\infty d\tilde{b}\nonumber\\
	&\hspace{1em}\times \tilde{b}c_{12}^7 \sin^2\chi^{(0)}\exp\left(-\frac{\varepsilon}{2T}c_{12}^2\right),\label{eq:nu_0_1}\\
\nu^{(1)*}_1(T)&\equiv \frac{4}{15}\int_0^\infty dc_{12} \int_0^\infty d\tilde{b}
	\hspace{0.2em}\tilde{b}\left(\mathfrak{N}^2-\tilde{b}^2\right)c_{12}^7\nonumber\\
	&\hspace{1em}\times 
	\left(1-3\cos^2\frac{\chi^{(0)}}{2}\right)\exp\left(-\frac{\varepsilon}{2T}c_{12}^2\right),\label{eq:nu_1_1}\\
\nu^{(1)*}_2(T)&\equiv \frac{2}{5}\int_0^\infty dc_{12} \int_0^\infty d\tilde{b}\nonumber\\
	&\hspace{1em}\times \tilde{b}c_{12}^7\chi^{(1)}\sin2\chi^{(0)}\exp\left(-\frac{\varepsilon}{2T}c_{12}^2\right).\label{eq:nu_1_2}
\end{align}
Unfortunately, $\zeta(T)$ and $\nu(T)$ cannot be expressed explicitly. 
Therefore, we adopt the numerical integrals in Eqs.\ (\ref{eq:zeta})--(\ref{eq:nu_1_2}) over $\tilde{b}$ and $c_{12}$ for each $T/\varepsilon$. 

%%%%%%%%%%%%%%%%%%%%%%%%%%%%%%%%%%%%
\section{The normal stress differences}\label{sec:N1_N2}
%%%%%%%%%%%%%%%%%%%%%%%%%%%%%%
\begin{figure}[htbp]
	\begin{center}
		\includegraphics[width=80mm]{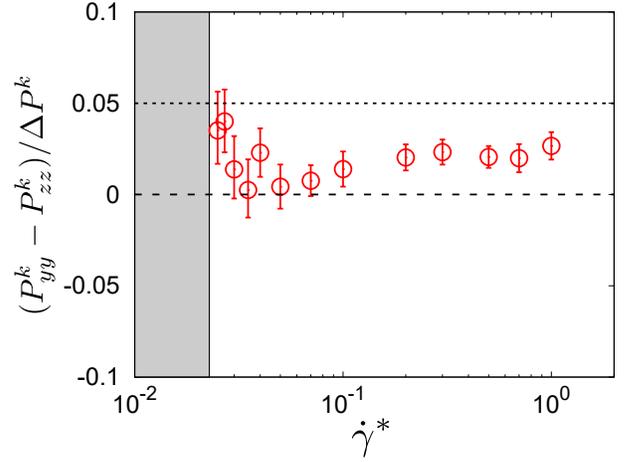}
	\end{center}
	\caption{The ratio of the second kinetic normal stress difference to the first kinetic normal stress difference $\Delta P$ for $e=0.99$.}
	\label{fig:N1N2}
\end{figure}
%%%%%%%%%%%%%%%%%%%%%%%%%%%%%%

In this Appendix, let us show that the second normal stress difference $P^k_{yy}-P^k_{zz}$ is much smaller than the first normal stress difference $\Delta P^k$ in our simulations.
Figure \ref{fig:N1N2} represents the shear rate dependence of the ratio of the second normal difference to the first normal stress difference obtained from the simulations, which shows that the ratio satisfies
\begin{equation}
\left|\frac{P^k_{yy}-P^k_{zz}}{\Delta P^k}\right|<0.05.
\end{equation}
This result also validates our treatment that we only consider the set of $p^k$, $P^k_{xy}$, and $\Delta P^k$ in our theoretical treatment.

%%%%%%%%%%%%%%%%%%%%%%%%%%%%%%%%%%%%
\section{Linear stability analysis}\label{sec:stability}
\begin{figure}[htbp]
	\begin{center}
		\includegraphics[width=80mm]{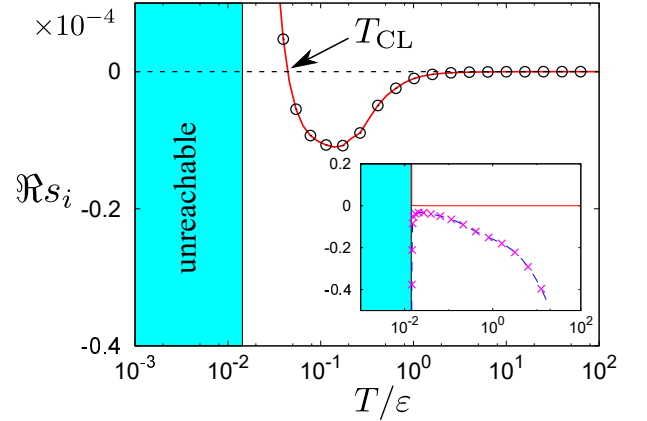}
	\end{center}
	\caption{The temperature dependence of the real part of the eigenvalue whose absolute value is smallest (solid line).
	The open circles are the linear solution of the Eq.\ (\ref{eq:linear_eigen}).
	The inset also shows other two eigenvalues $s_2$ (dashed line) and $s_3$ (cross marks).
	The painted area exhibits an unreachable region, where there is no steady state.}
	\label{fig:eigen}
\end{figure}

In this Appendix, we study the linear stability of the uniform shear state without any spatial fluctuation \cite{Hayakawa2016}.
Let us consider the steady state of the dimensionless temperature $T^*$, the dimensionless pressure difference $\Delta P^*=\Delta P/(n\varepsilon)$, and the dimensionless shear stress $P_{xy}^*\equiv P_{xy}/(n\varepsilon)$.
We add a perturbation around the steady state as $\delta \phi \equiv (\delta T^*,\delta\Delta P^*, \delta P_{xy}^*)$.
Introducing the dimensionless quantities $\dot\gamma^*\equiv \dot\gamma t_0$, $\zeta^*=\zeta t_0$ and $\nu^*\equiv \nu t_0$ with $t_0\equiv \sqrt{md^2/\varepsilon}$, the time evolution of the fluctuation is linearized as
\begin{equation}
	\frac{\partial}{\partial t^*} \delta \phi = {\cal M}\delta \phi,\label{eq:phi_evol}
\end{equation}
where the matrix ${\cal M}$ is defined by
\begin{align}
	&{\cal M}=\nonumber\\
	&\begin{pmatrix}
		\displaystyle
		-\left(\frac{2}{3}\dot\gamma^*_T P_{xy}^*+\zeta^*+\zeta^*_T T\right)
		& 0 & \displaystyle-\frac{2}{3}\dot\gamma^*\\
		\displaystyle-\left(2\dot\gamma^*_T P^*_{xy} +\nu^*_{T}\Delta P^*\right)
		& -\nu^* & -2\dot\gamma^*\\
		\displaystyle-\left(\dot\gamma^*+\dot\gamma^*_T T^*-\frac{1}{3}\dot\gamma^*_T\Delta P^*
		+\nu^*_{T}P^*_{xy}\right) & \displaystyle\frac{1}{3}\dot\gamma^* & -\nu^*
	\end{pmatrix}.
\end{align}
with $\dot\gamma^*+\dot\gamma^*_T \delta T^*$, $\zeta^*+\zeta^*_T \delta T^*$, and $\nu^*+\nu^*_{T}\delta T^*$ with $\dot\gamma^*_T\equiv (\partial \dot\gamma^*/\partial T^*)_T$, 
$\zeta^*_T\equiv (\partial \zeta^*/\partial T^*)_T$, and $\nu^*_{T}\equiv (\partial \nu^*/\partial T^*)_T$.
The Laplace transform of Eq.\ (\ref{eq:phi_evol}) is expressed as
\begin{equation}
	\varphi(s)=(s\overleftrightarrow{1}-{\cal M})^{-1}\delta \phi(0),
\end{equation}
where $\varphi(s)={\cal L}[\delta \phi(t)]$ is the Laplace transform of $\delta \phi(t)$.
Let us assume $s_i$ ($i=1,2,3$) as the eigenvalues of the matrix ${\cal M}$.
This yields that the matrices $(s\overleftrightarrow{1}-{\cal M})$ and $(s\overleftrightarrow{1}-{\cal M})^{-1}$ have eigenvalues $s-s_i$ and $1/(s-s_i)$ ($i =1,2,3$), respectively. 
The inverse Laplace transform of $1/(s-s_i)$ is given by $\exp(s_i t)\Theta(t)$, where $\Theta(t)$ is the step function, i.e., $\Theta(t)=1$ for $t\ge 0$ and $\Theta(t)=0$ otherwise.
The system becomes unstable if any one of the eigenvalues have positive real part. 
Let us calculate the eigenvalues numerically.
The determinant of $s_i\overleftrightarrow{1}-{\cal M}$ is given by
\begin{equation}
\det (s_i \overleftrightarrow{1}-{\cal M})=s_i^3+A_1 s_i^2+A_2s_i + A_3=0,\label{eq:eigen}
\end{equation}
where $A_1$, $A_2$, and $A_3$ are, respectively, given by
\begin{align}
A_1&=\frac{2}{3}\dot\gamma^*_T P_{xy}^* +\zeta^*+\zeta^*_T T^*+2\nu^*,
\end{align}
%%%%%%%%%%%%%%%%%%%%%%%%%%%%%%	
\begin{align}
A_2&=\nu^{*2}+2\nu^*\left(\frac{2}{3}\dot\gamma^*_T P_{xy}^* 
	+\zeta^*+\zeta^*_T T^*\right)\nonumber\\
	&\hspace{1em}
	-\frac{2}{3}\dot\gamma^*\left(\dot\gamma^*_T T^*-\frac{1}{3}\dot\gamma^* \Delta P^*
	+\nu^*_T P^*_{xy}\right),\\
A_3 &= \left(\frac{2}{3}\dot\gamma^{*2}+\nu^{*2}\right)
	\left(\frac{2}{3}\dot\gamma_T^*P_{xy}^* + \zeta^* +\zeta^*_T T^*\right)\nonumber\\
	&\hspace{1em}
	-\frac{2}{9}\dot\gamma^{*2}(2\dot\gamma^*_T P_{xy}^*+\nu_T^*\Delta P^*)\nonumber\\
	&\hspace{1em}
	-\frac{2}{3}\nu^*\dot\gamma^*
	\left(\dot\gamma^*+\dot\gamma^*_T T^*-\frac{1}{3}\dot\gamma^*_T\Delta P^*
	+\nu^*P_{xy}^*\right).
\end{align}
We also focus on the eigenvalue whose absolute value is smallest.
Neglecting the terms proportional to $s_i^2$ and $s_i^3$ in Eq.\ (\ref{eq:eigen}), we can obtain the linear approximation solution of Eq.\ (\ref{eq:eigen}) as
\begin{equation}
	s_{\rm linear} = -\frac{A_3}{A_2}.\label{eq:linear_eigen}
\end{equation}
Figure \ref{fig:eigen} presents the temperature dependence of the real part of each eigenvalue $s_i$ ($\Re s_1 > \Re s_2\ge \Re s_3$), where $\Re s_i$ indicates the real part of $s_i$.
We also plot the linear approximation solution (\ref{eq:linear_eigen}).
This linearized eigenvalue in Eq.\ (\ref{eq:linear_eigen}) gives a good description of the full linear stability analysis.
Note that the real part of eigenvalue becomes positive for $T<T_{\rm CL}\equiv 0.04\varepsilon$ in which the steady state is unstable.
Near the critical temperature, the magnitude of the smallest eigenvalue is approximately $10^{-5}$.
We note that approximately $10^5$ collisions per particle are needed to reach the steady state in this regime.
%%%%%%%%%%%%%%%%%%%%%%%%%%%%%%
%%%%%%%%%%%%%%%%%%%%%%%%%%%%%%

%%%%%%%%%%%%%%%%%%%%%%%%%
\section{Critical behavior}\label{sec:critical_temp}
\begin{figure}
	\begin{center}
		\includegraphics[width=80mm]{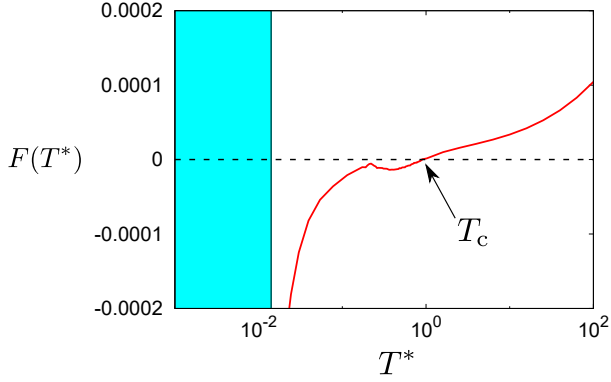}
	\end{center}
	\caption{The temperature dependence of Eq.\ (\ref{eq:Tc3}) for $e=0.99$.
	The shaded area exhibits an unreachable region, where there is no steady state.}
	\label{fig:Tc3}
\end{figure}

In this Appendix, we calculate the critical temperature for the linear stability analysis, where $\partial \dot\gamma/\partial T$ becomes zero.
From Eq.\ (\ref{eq:steady_gamma}), $\partial \dot\gamma/\partial T$ satisfies
\begin{align}
2\dot\gamma \frac{\partial \dot\gamma}{\partial T}
&= \frac{3}{2}\frac{\nu}{(\nu-\zeta)^2}\left[(\nu-2\zeta)\zeta\frac{\partial \nu}{\partial T}+\nu^2\frac{\partial \zeta}{\partial T}\right].
\end{align}
This means that $\nu=0$ or
\begin{equation}
(\nu-2\zeta)\zeta\frac{\partial \nu}{\partial T}+\nu^2\frac{\partial \zeta}{\partial T}=0,\label{eq:condition2}
\end{equation}
should be satisfied at the temperature satisfying $\partial \dot\gamma/\partial T=0$. 
As shown in Fig.\ \ref{fig:nu}, $\nu$ becomes zero at $T\simeq 0.00451\varepsilon$.
Let us calculate other temperatures where $\partial \dot\gamma/\partial T$ becomes zero. 
Using the dimensionless quantities, the left hand side of Eq.\ (\ref{eq:condition2}) can be rewritten as
\begin{align}
F(T^*)&=(\nu^*-2\zeta^*)\zeta^*\frac{\partial \nu^*}{\partial T^*}+\nu^{*2}\frac{\partial \zeta^*}{\partial T^*}.\label{eq:Tc3}
\end{align}
Figure \ref{fig:Tc3} shows that Eq.\ (\ref{eq:Tc3}) has only one solution $T_{\rm c}=0.910\varepsilon$.
The corresponding shear rate becomes $\dot\gamma_{\rm c}^*=0.0228$ and any steady state does not exist for $\dot\gamma<\dot\gamma_{\rm c}$.
%%%%%%%%%%%%%%%%%%%%%%%%%%%%%%
%%%%%%%%%%%%%%%%%%%%%%%%%%%%%%

Let us expand the shear viscosity around the critical temperature $T_{\rm c}$.
First, the quantities $\nu$ and $\zeta$ are expanded as
\begin{align}
	\nu &= \nu_{\rm c}+\nu_{\rm c}^\prime (T-T_{\rm c}) 
		+\frac{1}{2}\nu_{\rm c}^\pp (T-T_{\rm c})^2+ \mathcal{O}\left((T-T_{\rm c})^3\right),\\
	\zeta &= \zeta_{\rm c}+\zeta_{\rm c}^\prime (T-T_{\rm c})
		+\frac{1}{2}\zeta_{\rm c}^\pp (T-T_{\rm c})^2+ \mathcal{O}\left((T-T_{\rm c})^3\right),
\end{align}
respectively, where $\nu_{\rm c}=\nu(T_{\rm c})$, $\nu_{\rm c}^\prime=(d\nu/dT)_{T_{\rm c}}$, $\nu_{\rm c}^\pp=(d^2\nu/dT^2)_{T_{\rm c}}$, $\zeta_{\rm c}=\zeta(T_{\rm c})$, $\zeta_{\rm c}^\prime=(d\zeta/dT)_{T_{\rm c}}$, and $\zeta_{\rm c}^\pp=(d^2\zeta/dT^2)_{T_{\rm c}}$, respectively.
Inserting into Eq.\ (\ref{eq:steady_gamma}) and Eq.\ (\ref{eq:steady_eta}), the shear rate and the shear viscosity are, respectively, expressed as
\begin{align}
	\frac{\dot\gamma}{\dot\gamma_{\rm c}}
	&=1+C_1(T-T_{\rm c})^2+\mathcal{O}\left((T-T_{\rm c})^3\right),\\
	\frac{\eta}{\eta_{\rm c}}&= 1+C_2 (T-T_{\rm c})+\mathcal{O}\left((T-T_{\rm c})^2\right),
\end{align}
where $\dot\gamma_{\rm c}$, $\eta_{\rm c}$, $C_1$ and $C_2$ are, respectively, given by
\begin{align}
	\dot\gamma_{\rm c}&= \sqrt{\frac{3}{2}\frac{\nu_{\rm c}^2\zeta_{\rm c}}{\nu_{\rm c}-\zeta_{\rm c}}},\label{eq:gamma_c}\\
	\eta_{\rm c}&=\frac{\nu_{\rm c}-\zeta_{\rm c}}{\nu_{\rm c}^2}nT_{\rm c}^\prime,\\
	C_1 &=\frac{\zeta_{\rm c}(\nu_{\rm c}-2\zeta_{\rm c})\nu_{\rm c}^\pp +\nu_{\rm c}^2\zeta_{\rm c}^\pp}
		{4\nu_{\rm c}\zeta_{\rm c}(\nu_{\rm c}-\zeta_{\rm c})}\nonumber\\
		&\hspace{1em}-\frac{(\nu_{{\rm c}}-4\zeta_{\rm c})(\zeta_{\rm c}\nu_{\rm c}^\prime - \nu_{\rm c}\zeta_{\rm c}^\prime)^2}
		{8\nu_{\rm c}\zeta_{\rm c}^2(\nu_{\rm c}-\zeta_{\rm c})^2},\\
	C_2 &= \frac{1}{T_{\rm c}} - \frac{(\nu_{\rm c}-2\zeta_{\rm c})\nu_{\rm c}^\prime+\nu_{\rm c}\zeta_{\rm c}^\prime}{\nu_{\rm c}(\nu_{\rm c}-\zeta_{\rm c})}.
\end{align}
In the vicinity of the critical temperature $T_{\rm c}$, the relationship between the shear viscosity and the shear rate becomes
\begin{equation}
	\eta-\eta_{\rm c}=\pm \sqrt{\frac{C_2^2}{C_1}\frac{\eta_{\rm c}^2}{\dot\gamma_{\rm c}}} \left(\dot\gamma-\dot\gamma_{\rm c}\right)^{1/2}.\label{eq:eta_c_gamma_c}
\end{equation}
Figure \ref{fig:steady_gamma_eta} shows this expansion with well agrees with both the results of our simulation and the relationship obtained by Eqs.\ (\ref{eq:steady_gamma}) and (\ref{eq:steady_eta}).

%%%%%%%%%%%%%%%%%%%%%%%%%%%%%%
%%%%%%%%%%%%%%%%%%%%%%%%%%%%%%

\end{document}